\documentclass[reprint,superscriptaddress,pre,twocolumn,amsmath,amssymb,aps,footinbib]{revtex4-2}

\usepackage{graphics}
\usepackage{graphicx}
\usepackage{xr-hyper}
\usepackage{hyperref}
\usepackage{xcolor}
\hypersetup{colorlinks=true, linkcolor=blue, citecolor=blue, urlcolor=blue, filecolor=blue}
\usepackage{lipsum}
\usepackage[caption=false]{subfig}
\usepackage{mathtools}
\usepackage{enumerate}
\usepackage{tikz}
\usepackage{array}

\DeclareRobustCommand{\alignarrow}{
\tikz[baseline=-0.2ex, x=0.75pt, y=0.75pt, yscale=1, xscale=1, scale=0.02]{
\useasboundingbox (300,7) rectangle (800,245);
\draw[fill=black, fill opacity=1] (9,108.52) -- (445.5,108.52) -- (445.5,7) -- (644.5,126) -- (445.5,245) -- (445.5,143.48) -- (9,143.48) -- cycle;
}}

\DeclareRobustCommand{\antialignarrow}{%
\tikz[baseline=-2.25ex, x=0.75pt, y=0.75pt, yscale=-1, xscale=1, scale=0.06]{%
\useasboundingbox (25,20) rectangle (220,120);  
\draw[fill=black, fill opacity=1] (172.5,108) -- (86.5,107.77) -- (50.54,67.8) -- (86.71,28.01) -- (172.71,28.23) -- (136.54,68.02) -- cycle;
\draw[draw opacity=0, fill=white, fill opacity=1] (101.23,21) -- (100.5,118) -- (45.11,117.58) -- (45.84,20.58) -- cycle;
\draw[line width=1] (115.5,164) .. controls (-5.5,202) and (1.5,39) .. (111.5,67);
}}

\usepackage[american]{babel}

\newcommand{\rbf}{\mathbf{r}}
\newcommand{\pbf}{\hat{\mathbf{p}}}
\newcommand{\kbf}{\mathbf{k}}

\newcommand{\RI}{\sigma}

\newcommand\underrel[3][]{\mathrel{\mathop{#3}\limits_{%
			\ifx c#1\relax\mathclap{#2}\else#2\fi}}}

\renewcommand{\Re}{\operatorname{Re}}
\renewcommand{\Im}{\operatorname{Im}}

\newcommand{\del}{\partial}

\renewcommand{\epsilon}{\varepsilon}

\newcommand{\gAA}{J_{AA}}
\newcommand{\gAB}{J_{AB}}

\newcommand{\rholsa}{\rho}

\begin{document}

\title{Kinetic theory of pattern formation in a generalized multispecies Vicsek model}

\date{\today}

\begin{abstract}
	\noindent
The theoretical understanding of pattern formation in active systems remains a central problem of interest. Heterogeneous flocks made up of multiple species can exhibit a remarkable diversity of collective states that cannot be obtained from single-species models. In this paper, we derive a kinetic theory for multispecies systems of self-propelled particles with (anti)alignment interactions. We summarize the numerical results for the binary system before employing linear stability analysis on the coarse-grained system. We find good agreement between theoretical predictions and particle simulations, and our kinetic theory is able to capture the correct lengthscale in the emergent coexistence phases through a Turing-Hopf instability. Extending the kinetic framework to multispecies systems with cyclic alignment interactions, we recover precisely the same emergent ordering as corresponding simulations of the microscopic model. More generally, our kinetic theory provides an extensible framework for analyzing pattern formation and collective order in multispecies active matter systems.
\end{abstract}

\author{Eloise Lardet}
\author{Letian Chen}
\author{Thibault Bertrand}
\email[Electronic address: ]{t.bertrand@imperial.ac.uk}
\affiliation{Department of Mathematics, Imperial College London,
180 Queen’s Gate, London SW7 2AZ, United Kingdom}

\maketitle


\section{Introduction}

Far from equilibrium, order emerges in unexpected ways when alignment gives way to antialignment. The emergence of order in nonequilibrium systems indeed challenges traditional notions of phase transitions in equilibrium settings, and even more so when multiple species interact. There, even simple alignment or antialignment rules can give way to a collection of patterns. In single-species active systems, simple interaction rules are already sufficient to produce a variety of nontrivial collective behaviors \cite{ramaswamy2010,menzel2015,bechinger2016}. In this context, theoretical explorations have been strongly motivated by the remarkable patterns observed in natural systems. From collective motion in animal groups \cite{giardina2008,cavagna2014}, to bacterial clustering \cite{peruani2012}, and ordered states in myosin filaments \cite{schaller2010,huber2018}, nature never ceases to astound and inspire physicists.

Polar and nematic flocking are classic examples of collective behavior in active matter, with much of their theoretical foundation built on the analysis of the now classical Vicsek model \cite{vicsek1995, gregoire2003, gregoire2004,chate2006,chate2008a,peruani2011,yuan2024}. In the Vicsek model, self-propelled particles experience local alignment interactions leading to a transition from disorder to ordered flocks \cite{vicsek2012,chate2020}. Nonetheless, flocking has been documented in a broad spectrum of systems and shown to arise even with modified alignment rules (e.g. metric-free interactions, self-aligning interactions or even pure antialignment interactions) \cite{ginelli2010,knezevic2022,das2024,baconnier2025} and even in the absence of explicit alignment mechanisms \cite{romanczuk2009,deseigne2012,casiulis2022,caprini2023}. Furthermore, particles often self-organize in surprising ways in the presence of repulsion \cite{grossmann2013}, volume effects \cite{martin-gomez2018,sese-sansa2018}, or to relieve frustration generated by antialigning interactions \cite{grossmann2015,denk2020,lardet2024,escaff2024a,escaff2024b,lardet2026}.

The rich behavior of active systems, in both theoretical and experimental studies, calls for analytical approaches that can bridge microscopic dynamics and macroscopic phenomena \cite{toner2005, marchetti2013}. Various approaches have been used to derive hydrodynamic theories of Vicsek-like models. While Toner and Tu pioneered the hydrodynamic description of polar flocks using a top-down symmetry-based approach \cite{toner1995,toner1998}, the precise nature of the transition to flocking remains a subject of intense theoretical scrutiny as their original conclusions were later put in question \cite{toner2012}. Promising progress was made recently by Jentsch and Lee \cite{jentsch2024}, who reported exact predictions agreeing with high-precision numerical studies of the Vicsek model \cite{mahault2019}. So did Chaté and Solon who proposed simultaneously exact scaling relations also demonstrating close agreement with these numerical results \cite{chate2024}. In response to this claim, Chen {\it et al.} argued that neglected nonlinear couplings in number-conserving systems prohibit such exactness, while Chaté and Solon maintain that these couplings are symmetry-forbidden \cite{chate2024,chen2025a,chate2025a,chen2025b,chate2025b}. 

Other strategies follow instead a bottom-up approach, starting from microscopic equations to derive hydrodynamic equations through kinetic theories and coarse-graining approaches. Such strategies allow one to establish the form of the hydrodynamic equations but also determine the dependence of transport coefficients on physical parameters. For instance, the Boltzmann-Ginzburg-Landau (BGL) approach, rooted in the Boltzmann equation and based on a low-density binary collision assumption, has successfully been used to describe the disorder-to-order transition in the Vicsek model \cite{bertin2006, bertin2009, peshkov2014} and nematic systems \cite{bertin2013}. Alternatively, the Smoluchowski approach, often based on a mean-field Fokker-Planck formulation, has been used to derive continuum level hydrodynamic equations for Vicsek-like models to understand the large-scale dynamics and spatial organization, bridging continuum theory to collective phenomena \cite{farrell2012,marchetti2013,grossmann2013}. Boltzmann-Ginzburg-Landau (BGL) and Smoluchowski schemes have been systematically compared, demonstrating that while both yield identical hydrodynamic structures for simple point-like particles, significant discrepancies arise for self-propelled rods \cite{bertin2015}. They concluded that the BGL framework offers better control, as it naturally resolves ambiguities regarding interaction terms that are often obscured by the closure approximations required in the Smoluchowski formalism \cite{bertin2015}. However, as the BGL approach is limited to systems with tractable interaction kernels, Smoluchowski approaches are often favored for complex extensions of Vicsek-like models. We note that alternative kinetic theories of Vicsek-like systems also exist \cite{ihle2011,degond2013}. Importantly, all these methods involve approximations; the choice of which to employ depends subtly on the specific model and the phenomena one wishes to capture.

In recent years, interest has shifted beyond single-species models to binary mixtures of self-propelled particles \cite{menzel2012,chatterjee2023,maity2023}. In these systems, antialignment interactions have been shown to produce novel coexistence patterns \cite{kursten2025, oki2026, lardet2026}, while the introduction of nonreciprocal couplings has opened a new frontier in nonequilibrium physics and attracted significant attention \cite{you2020,saha2020,fruchart2021,kreienkamp2024a,kreienkamp2024b,mangeat2025,myin2026}. Moving to general multispecies architectures, there exists a vast literature on cyclic interaction networks (e.g., Rock-Paper-Scissors games) used for instance to study competition in biological and social systems \cite{szolnoki2020}. While these topologies are standard in population dynamics, they have seldom been studied in the context of active matter systems. Notably, systems driven by nonreciprocal attraction-repulsion cycles have revealed remarkable self-organized dynamics, including chasing phases and active molecules \cite{ouazan-reboul2023,ouazan-reboul2023a,ouazan-reboul2023b}.

In Ref.\,\cite{lardet2026}, we generalized the two-species Vicsek model to include arbitrary reciprocal intra- and interspecies couplings. Using extensive particle-based simulations, we uncovered a rich phenomenology where, counter-intuitively, antialignment interactions can drive phase separation and global polar order, leading to stable periodic traveling bands. We further extended this to multispecies ($m>2$) systems, demonstrating that cyclic alignment interactions can generate stable traveling stripes whose behavior depends critically on the parity of the cycle length, analogous to the scalar dynamics observed in nonreciprocal attraction-repulsion cycles \cite{ouazan-reboul2023,ouazan-reboul2023a,ouazan-reboul2023b}. In this paper, we complement these numerical findings by deriving a comprehensive kinetic theory based on a Smoluchowski approach, allowing us to analytically explain the origin and stability of the coexistence phases reported in \cite{lardet2026}.

The paper is organized as follows. In Sec.~\ref{sec:model}, we introduce the microscopic model. Section \ref{sec:kinetic_theory} presents a kinetic theory of the coarse-grained system, derived via an angular Fourier mode analysis of the Fokker-Planck equation. In Sec.~\ref{sec:two_species_results}, we apply linear stability analysis to the two-species system, identifying a region of finite-wavelength instability and comparing these predictions with particle-based simulations. Section \ref{sec:mutli_species_results} extends this framework to multispecies systems with cyclic alignment, demonstrating how our theory rationalizes the emergent ordering observed in simulations. Finally, we provide concluding remarks in Sec.~\ref{sec:conclusion}. Technical details on the numerical methods and stability analysis are provided in the Appendices.

\section{Model} \label{sec:model}

We consider a two-dimensional model of $N$ self-propelled point particles belonging to $m$ species. We assume an equal proportion of species such that $N = m {\cal M}$, where ${\cal M}$ is the number of particles in each species. Particles are confined to a two-dimensional periodic domain of size $L\times L$ with a global number density of $\rho=N/L^2$. The state of each particle $i$ is defined by its position $\mathbf{r}_i$ and orientation angle $\theta_i$, with a unit heading vector $\pbf_i=(\cos\theta_i,\sin\theta_i)^T$. Particles move with self-propulsion speed $v_0$ in their heading direction. They experience rotational diffusion and interact through a local alignment torque (metric interactions). The position and orientation of each particle $i$ are described by the following overdamped Langevin equations:
\begin{subequations}
	\begin{align}
		\dot{\mathbf{r}}_i &= v_0 \pbf_i, \label{eq:langevin_r} \\
		\dot{\theta}_i &= \sum_{j\in\mathcal{N}_i} J_{s(i), s(j)} \sin(\theta_j-\theta_i) + \sqrt{2D_r} \ \xi_i, \label{eq:langevin_theta}
	\end{align} \label{eq:langevin}%
\end{subequations}
where $\xi_i(t)$ is a zero-mean, unit-variance Gaussian white noise. The first term in Eq.\,(\ref{eq:langevin_theta}) is the local alignment torque, where the summation runs over the set of neighbors $\mathcal{N}_i\equiv\{j : |\mathbf{r}_i-\mathbf{r}_j|\leq \sigma, i\neq j\}$. The coupling constant $J_{s(i), s(j)}$ is determined by the species indices $s(i)$ and $s(j)$, and results in alignment (ferromagnetic) interactions if $J_{s(i),s(j)}>0$ or antialignment (antiferromagnetic) interactions if $J_{s(i),s(j)}<0$. 
We emphasize that Eq.\,(\ref{eq:langevin_theta}) employs an additive interaction rule (unnormalized sum). 
While standard Vicsek-like simulations often use a mean-sine model with a normalization of the alignment term by the number of neighbors $|\mathcal{N}_i|$, this additive form significantly simplifies the analytical treatment while capturing all the relevant phases and phase co-existences reported in \cite{lardet2026}.

For our particle-based simulations, we numerically solve Eq.\,(\ref{eq:langevin}) using an Euler-Maruyama scheme with timestep $\Delta t=5\times10^{-3}$, self-propulsion speed $v_0=1$ and interaction range $\sigma=1$ (unless stated otherwise), starting from a disordered homogeneous state.

\section{Kinetic theory} \label{sec:kinetic_theory}

\subsection{Smoluchowski approach}

Starting from the microscopic Langevin equations (\ref{eq:langevin}), we use Dean's approach to obtain a Fokker-Planck equation for the one-particle probability distribution \cite{dean1996}.
To do so, we extend the procedure followed by Refs.~\cite{grossmann2015} and \cite{escaff2024a} to multiple species similarly to what was done in Ref.~\cite{fruchart2021}. 

We write the particle density as 
\begin{align}
	c^a(\rbf, \theta, t) = \sum_{i=1}^{{\cal M}}\delta(\rbf-\rbf_i^a(t))\delta(\theta-\theta_i^a(t)),
\end{align}
for each species $a\in\{1, \cdots, m\}$. Upon a change of variables, using It\^o calculus and neglecting stochastic fluctuations (valid in the high density case \cite{grossmann2014}), we obtain the following general multispecies Fokker-Planck equation
\begin{equation} \label{eq:FPE}
	\del_t c^a = -v_0 \pbf \cdot \nabla_{\rbf} c^a + D_r \del_\theta^2 c^a - \del_\theta \mathcal{T}[c^a],
\end{equation}
with
\begin{align}
	\mathcal{T}[c^a] = c^a&(\rbf, \theta, t) \times  \\ 
								& \int\limits_{{\cal B}_\sigma(\mathbf{r})} d^2\mathbf{r}' \int\limits_0^{2\pi} d\theta' \sum_b J_{ab} c^b(\rbf', \theta', t)\sin(\theta'-\theta), \nonumber
\end{align}
where ${\cal B}_{\RI}(\rbf)$ is the ball of radius $\RI$ centered at point $\rbf$.


\subsection{Fourier expansion and linear stability analysis}

To make progress, we move to Fourier space to analyze Eq.\,(\ref{eq:FPE}). Defining the Fourier transform as
\begin{equation}
	f_n^a(\rbf,t) = \int d\theta e^{in\theta} c^a(\rbf, \theta, t),
\end{equation}
we can express the particle density in terms of its angular mode decomposition as follows 
\begin{equation}
	c^a(\rbf,\theta,t) = \frac{1}{2\pi} \sum_n e^{-in\theta} f_n^a(\rbf,t).
\end{equation}
Upon expansion, re-indexing and performing all angular derivatives, we obtain at order $n$ the following governing equation for the angular mode $f_n^a(\mathbf{r},t)$
\begin{widetext}
\begin{align}
	 \del_t f_n^a(\rbf, t) = &-\frac{v_0}{2} \Big[\del_zf_{n+1}^a(\rbf,t) + \del_{\bar{z}} f_{n-1}^a(\rbf,t)\Big] - D_r n^2 f_n^a(\rbf,t) \nonumber \\
	 &+ \sum_b \frac{J_{ab}}{4\pi} \int\limits_{B_{\sigma}(\rbf)} d^2 \mathbf{r}' \int\limits_0^{2\pi} d\theta' \sum_{n'} n  e^{-in'\theta'} \Big(f_{n-1}^a(\rbf,t) f_{n'+1}^b(\rbf',t) - f_{n+1}^a(\rbf,t) f_{n'-1}^b(\rbf',t)\Big),
\end{align}
where we have used that
\begin{equation}
	\pbf (\theta) \cdot \nabla_{\rbf} = \frac{1}{2}(e^{i\theta}\del_z + e^{-i\theta}\del_{\bar{z}}),
\end{equation}
and defined
\begin{equation}
	\del_z \equiv \del_x - i\del_y, \qquad \del_{\bar{z}} \equiv \del_x + i \del_y.
\end{equation}
Noting that $\int_0^{2\pi} d\theta e^{-in\theta} = 2\pi \delta_{n,0}$, where $\delta_{n,0}$ is the Kronecker delta function, we have
	\begin{align} \label{eq:angular_expansion}
	 \del_t f_n^a(\rbf,t) = & -\frac{v_0}{2} \Big[\del_z f_{n+1}^a(\rbf,t) + \del_{\bar{z}} f_{n-1}^a(\rbf,t)\Big] - D_r n^2 f_n^a(\rbf,t) \nonumber \\
	 				&+ \frac{n}{2}\sum_b J_{ab} \int\limits_{B_{\sigma}(\rbf)} d^2 \mathbf{r}' \Bigl[f_{n-1}^a(\rbf,t) f_{1}^b(\rbf',t) - f_{n+1}^a(\rbf,t) f_{-1}^b(\rbf',t)\Bigr].
\end{align}
Rather than performing a closure operation by letting $\del_t f^a_2=0$ and $f^a_3=0$, as commonly done in the literature \cite{bertin2006,farrell2012,fruchart2021}, we keep all orders of $n$ for now (as in Refs.~\cite{grossmann2013,grossmann2015,escaff2024a}). Furthermore, we treat the interaction kernel exactly, rather than approximating it by a delta function as done in Refs.~\cite{fruchart2021,marchetti2013}.
\end{widetext}

\subsubsection{Perturbation around the disordered state}

The Fokker-Planck equation [Eq.\,(\ref{eq:FPE})] always has a trivial uniform disordered solution
\begin{equation}
c^a_0= \rholsa^a/2\pi, \quad 1 \le a \le m 
\end{equation}
where $\rholsa^a$ is the system density of species $a$. We perturb around the disordered solution and write our infinitesimal perturbation as a plane-wave mode such that
\begin{equation}
	c^a(\rbf,\theta,t) = \frac{\rholsa^a}{2\pi} + \epsilon e^{\mu(k) t + i \kbf \cdot \rbf} \Phi^a(\theta),
\end{equation}
where $\epsilon \ll 1$ is a small perturbation amplitude, $\kbf=(k_x,k_y)$ is the wave vector with magnitude $k\equiv|\kbf|$ and $\Phi^a(\theta)$ is the angular dependence of the perturbation. Thus, following the angular Fourier transform, we have 
\begin{equation} \label{eq:perturbation_fourier}
	f^a_n(\rbf, t) = \rholsa^a \delta_{n,0} + \epsilon e^{\mu(k) t + i \kbf \cdot \rbf} \Phi_n^a.
\end{equation}

We insert this perturbation into Eq.\,(\ref{eq:angular_expansion}) and linearize with respect to $\epsilon$. Most of the terms are straightforward to deal with, apart from the interaction term, which requires more attention. Letting this integral interaction term be called $K_\textrm{int}$, inserting the perturbation Eq.\,(\ref{eq:perturbation_fourier}), and only looking at linear terms in $\epsilon$, we get
\begin{align}
	K_\textrm{int} = \frac{n}{2} & \sum_b J_{ab} \epsilon e^{\mu(k) t} \rholsa^a \times \\
	 & \int\limits_{B_{\RI}(\rbf)} d^2 \mathbf{r}'  e^{i\kbf \cdot \rbf'}\Bigl[ \delta_{n,1} \Phi_{1}^b - \ \delta_{n,-1}\Phi^b_{-1}  \Bigr]. \nonumber
\end{align}

We can calculate the integral of $e^{i\kbf\cdot\rbf}$ over the ball of radius $\RI$ using polar coordinates, finding that
\begin{equation} \label{eq:bessel_identity}
	\int\limits_{B_{\RI}(\rbf)} d^2 \mathbf{r}' e^{i\kbf \cdot \rbf'} = \frac{2\pi\RI}{k} J_1(\RI k)e^{i\kbf\cdot \rbf},
\end{equation}
with $J_1$ the Bessel function of the first kind of order 1. 

Using this identity, we obtain
\begin{align}
K_\textrm{int} =\epsilon e^{\mu t+i\kbf\cdot\rbf} \frac{\pi \RI}{k} J_1(\RI k) \rholsa^a \sum_b J_{ab} (\delta_{n,-1} + \delta_{n,1})\Phi_n^b .
\end{align}
Inserting the perturbation into the remaining terms, we arrive at an infinite hierarchy of equations of the form
\begin{equation} \label{eq:eigenvalue_problem}
	\mu(k) \Phi^a_n = \sum_b\sum_{l} (\mathcal{L}^{ab})_{nl} \Phi^b_l,
\end{equation}
\begin{widetext}
with the linear operator 
\begin{equation} \label{eq:lin_op_dis}
	(\mathcal{L}^{ab})_{nl} =  \delta_{a,b} \left[-\frac{iv_0}{2} \Bigl((k_x-ik_y)\delta_{n,l-1} + (k_x+ik_y) \delta_{n,l+1}\Bigr) - D_r n^2\delta_{n,l} \right] + \frac{\pi\RI}{k} \rholsa^a J_1(\RI k) J_{ab} (\delta_{n,-1} + \delta_{n,1}) \delta_{n,l}.
\end{equation}
\end{widetext}

Higher angular modes, corresponding to $|n|\gg 1$, are strongly damped as the diagonal $-D_r n^2$ terms then dominate; we can thus truncate these equations at a sufficiently large $|n|=N_c$ \cite{chou2012, grossmann2015}. This gives us an eigenvalue equation, with matrix $\mathcal{L}$, eigenvalues $\mu$ and eigenvector $\Phi$. We can express $\mathcal{L}$ as a block matrix. For example, for $m=2$ species, we have
\begin{equation} \label{eq:L_matrix_disordered}
	\mathcal{L} = 
	\begin{bmatrix}
		\mathcal{L}_{AA} & \mathcal{L}_{AB} \\
		\mathcal{L}_{BA} & \mathcal{L}_{BB}
	\end{bmatrix},
\end{equation}
where each block is of size $(2N_c+1) \times (2N_c+1)$.  The diagonal matrix blocks are tridiagonal, and the off-diagonal blocks have zeros everywhere except for the diagonal entries at $n=l=\pm1$. This matrix can become very large even for moderate values of $N_c$. We thus resort to computing the eigenvalues of this matrix numerically. 

\subsubsection{Perturbation around the ordered state}
We now look at perturbing around the spatially homogeneous ordered solution which takes the form of a von Mises distribution \cite{grossmann2013}
\begin{equation} \label{eq:von_mises}
	c^{a}_0(\theta) = \frac{\rholsa^a}{2\pi} \frac{\exp(\kappa^a  \cos(\theta-\theta_0^a))}{I_0(\kappa^a)},
\end{equation}
where $I_n(\cdot)$ is the modified Bessel function of the first kind of order $n$, $\theta_0^a$ is the direction of polar order, and $\kappa^a$ is an effective alignment parameter. Searching for a time-independent spatially uniform steady state solution to Eq.\,(\ref{eq:FPE}), we find that
\begin{equation} \label{eq:kappa}
	\kappa^a = \frac{\pi\sigma^2}{D_r}\sum_b J_{ab} \rholsa^b \Psi^b \cos(\theta_0^b - \theta_0^a),
\end{equation}
where $\Psi^b$ is the polar order parameter of species $b$ (see Appendix~\ref{sec:appendix_kappa} for a detailed derivation).

The order parameter is then determined self-consistently through the following transcendental equation \cite{grossmann2013}
\begin{equation} \label{eq:psi_bessel}
	\Psi^a = \frac{I_1(\kappa^a)}{I_0(\kappa^a)},
\end{equation}
in which $\kappa^a$ depends on the order parameters of all the species. We note that the trivial disordered solution $\Psi^a=0$, $1 \le a \le m$ always exists. 
However, for strong enough alignment and low noise, there exists a polar ordered solution, with order parameters $\Psi^a>0$. 

Having found an ordered solution (if one exists), then, we can once again perturb around this solution with a plane-wave mode. In Fourier space, this takes the form
\begin{equation} \label{eq:perturbation_fourier_ordered}
	f^a_n(\rbf, t) = \rholsa^a \frac{I^a_n}{I^a_0} e^{in\theta^a_0} + \epsilon e^{\mu(k) t + i \kbf \cdot \rbf} \Phi_n^a,
\end{equation}
where we have defined $I^a_n \equiv I_n (\kappa^a)$ for ease of notation. Inserting this perturbation into Eq.\,(\ref{eq:angular_expansion}) and linearizing we can once again derive an infinite hierarchy of equations. Special care is again required when considering the interaction integral, for which we eventually get
\begin{widetext}
\begin{align}
	K_\textrm{int} = & \ \epsilon e^{\mu t + i\kbf \cdot \rbf} n \sum_b J_{ab} \Biggl[ \rholsa^a \frac{\sigma\pi}{k} J_1(\sigma k) \left( \frac{I_{n-1}^a}{I_0^a}e^{i(n-1)\theta_0^a} \Phi_1^b - \frac{I_{n+1}^a}{I_0^a}e^{i(n+1)\theta_0^a}\Phi_{-1}^b \right) \nonumber \\
	& \hspace{4.75cm} + \rholsa^b \frac{\sigma^2\pi}{2}\left( \frac{I_1^b}{I_0^b}e^{i\theta_0^b} \Phi_{n-1}^a - \frac{I_{-1}^b}{I_0^b}e^{-i\theta_0^b} \Phi_{n+1}^a \right) \Biggr],
\end{align}
where we have used Eq.\,(\ref{eq:bessel_identity}) and the following result $\int\limits_{B_{\sigma}(\rbf)} d^2\mathbf{r}' = \pi \sigma^2$. This gives us a linear system of the form of Eq.\,(\ref{eq:eigenvalue_problem}). Noting that $I_{-1}=I_1$, we can finally write the linear operator in this case as
\begin{align} \label{eq:lin_op_ordered}
	(\mathcal{L}^{ab})_{nl} = \delta_{ab} & \left[ -\frac{iv_0}{2} \bigl((k_x-ik_y)\delta_{n,l-1} + (k_x+ik_y) \delta_{n,l+1}\bigr) - D_r n^2 \delta_{n,l} + \frac{n \pi \sigma^2}{2} \sum_c J_{ac} \rholsa^c \frac{I_1^c}{I_0^c} \Big(e^{i\theta_0^c}\delta_{n,l+1} - e^{-i\theta_0^c}\delta_{n,l-1}\Big) \right] \nonumber \\
	 & \
	+ J_{ab}\frac{n \pi \sigma}{k} J_1(\sigma k) \frac{\rholsa^a}{I_0^a} \left(I_{n-1}^a e^{i(n-1)\theta_0^a} \delta_{l,1} - I_{n+1}^a e^{i(n+1)\theta_0^a} \delta_{l,-1}\right).
\end{align}
\end{widetext}
From here, just as for the disordered state perturbation, we can truncate all modes $|n|$ larger than $N_c\gg1$ to form an eigenvalue equation which we solve numerically.

\section{Two-species systems} \label{sec:two_species_results}

In this section, we will study a generic system with two species ($m=2$). We will start by summarizing the phases and phase co-existences observed in our numerical simulations of the microscopic dynamics \cite{lardet2026}. We will then use the coarse-graining and linear stability analysis developed in Sec.\,\ref{sec:kinetic_theory} to study the stability of the homogeneous disordered and ordered phases observed. 

\begin{figure*}
	\includegraphics[width=\linewidth]{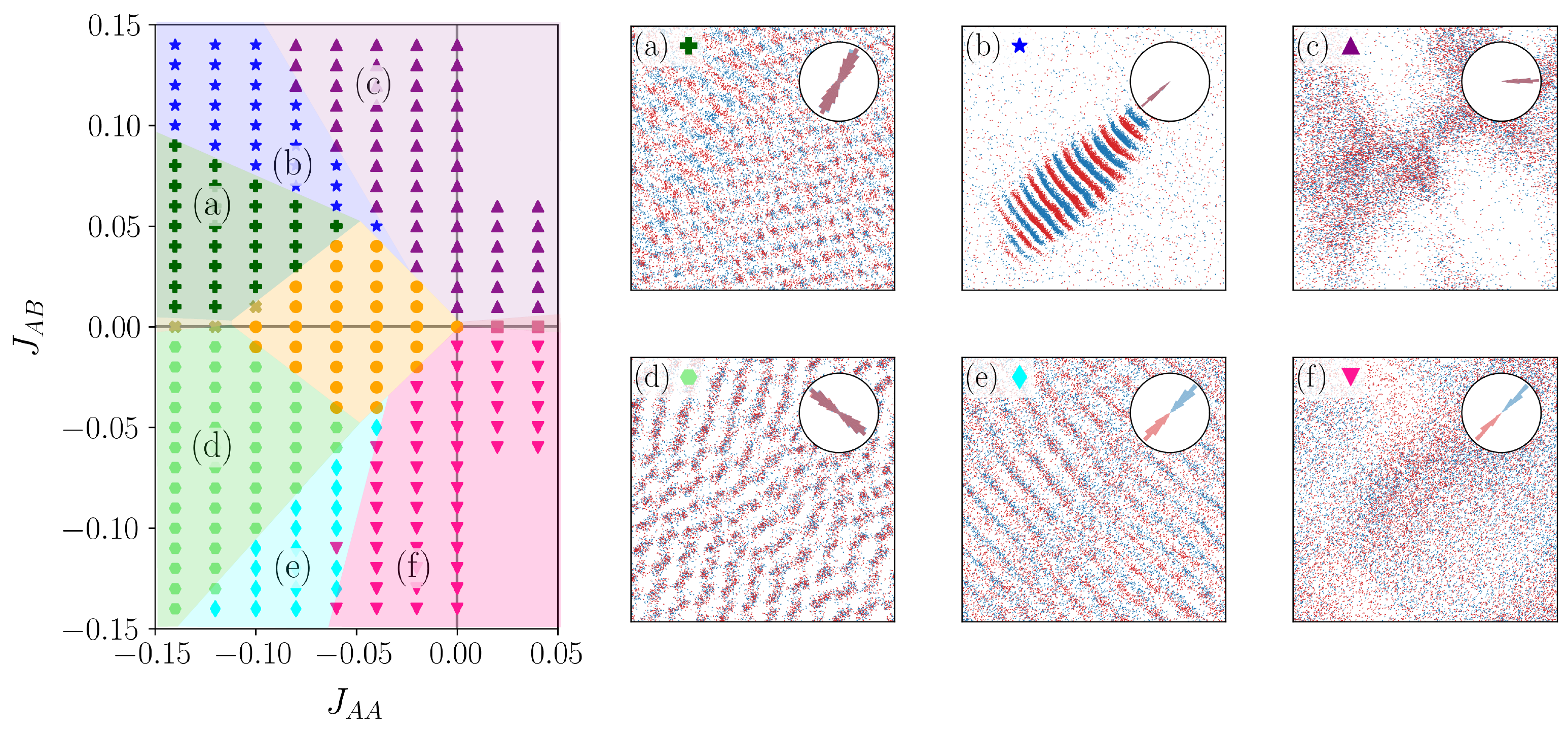}
	\caption{Phase diagram from particle simulations of the microscopic model with additive sine interactions. We solve Eq.\,(\ref{eq:langevin}) with the following parameters: $N=20000$, $\rho=100$, and $D_r=0.2$. Phases were classified numerically, with data averaged over 10 independent realizations for each data point. See details of the order parameters and classification in \cite{lardet2026} and Appendix \ref{sec:appendix_order_params}. Here, we list the phases observed and the parameters used to obtain the associated example snapshots: (a) Nematic stripes, demixed (dark-green plus), $\gAA=-0.12$, $\gAB=0.06$; (b) Flocking stripes (dark-blue star), $\gAA=-0.08$, $\gAB=0.08$; (c) Parallel flocking (purple upward triangle), $\gAA=-0.04$, $\gAB=0.12$; (d) Nematic stripes, mixed (light-green hexagon), $\gAA=-0.12$, $\gAB=-0.06$; (e) Antiparallel flocking stripes (light-blue diamond), $\gAA=-0.08$, $\gAB=-0.12$; (f)  Antiparallel flocking (pink downward triangle), $\gAA=-0.02$, $\gAB=-0.12$. We also observed a disordered state (orange circle), independent flocking (pink square), and independent nematic ordering (khaki cross). Phase boundaries were added manually as a visual guide.}
	\label{fig:phase_diagram_AS_microscopic}
\end{figure*}

\subsection{Simulation of microscopic equations} \label{sec:sim_m2}

We begin with the $m=2$ species case (i.e. binary systems), enforcing symmetric intraspecies couplings ($J_{AA}=J_{BB}$) and reciprocal interspecies couplings  ($J_{AB}=J_{BA}$) \footnote{The nonsymmetric parameter case is explored in the Supplemental Material of Ref.~\cite{lardet2026}.}. Although the phase diagram for mean-sine alignment was established in Ref.~\cite{lardet2026}, we verify here that all the phases persist in the additive sine model (Eq.~\ref{eq:langevin}) [see Fig.~\ref{fig:phase_diagram_AS_microscopic}]. We note that the additive nature of the interaction results in an effective rescaling of parameters by the average neighborhood size, $\langle|\mathcal{N}_i|\rangle\approx 300$.

Phases were identified via order parameters quantifying polar order, nematic order, demixing, and periodicity (see Appendix~\ref{sec:appendix_order_params} for details). First, we observe a liquid \textit{parallel flocking} phase for $\gAB>0$ and $\gAA>0$, which persists into the regime of weak intraspecies antialignment $\gAA<0$ [Fig.~\ref{fig:phase_diagram_AS_microscopic}(c)]. Analogously, a liquid \textit{antiparallel flocking} state emerges for $\gAB<0$ [Fig.~\ref{fig:phase_diagram_AS_microscopic}(f)], in which the species exhibit polar order in opposite directions. 
This phase and its transition from a disordered state in the case of fully antialigned couplings ($\gAA<0$, $\gAB<0$) was previously studied in Ref.~\cite{kursten2025}.

Increasing intraspecies antialignment triggers the emergence of striped phases. Near the boundary $\gAB\approx-\gAA$, a stable \textit{flocking stripes} phase appears [Fig.~\ref{fig:phase_diagram_AS_microscopic}(b)], featuring high-density bands of alternating species traveling in the same direction. Despite increased clustering compared to the mean-sine model \cite{lardet2026}, the characteristic wavelength remains $\lambda \approx 1.22$ (see Appendix~\ref{sec:appendix_stripe_wavelength}). Meanwhile, transitioning from liquid antiparallel flocking to stronger intraspecies antialignment ($\gAB\lesssim\gAA$) yields system-wide \textit{antiparallel flocking stripes} [Fig.~\ref{fig:phase_diagram_AS_microscopic}(e)], characterized by periodic bands of opposing polarity.

At stronger intraspecies antialignment, we observe nematically ordered coexistence patterns with spatially periodic bands or lanes parallel to the direction of motion. These are distinguished by their time-averaged demixing behavior: \textit{demixed nematic stripes} are encountered for $\gAB>0$ [Fig.~\ref{fig:phase_diagram_AS_microscopic}(a)] and \textit{mixed nematic stripes} are encountered for $\gAB<0$ [Fig.~\ref{fig:phase_diagram_AS_microscopic}(d)].

Finally, in the decoupling limit ($\gAB\approx0$), intraspecies interactions dominate, leading to independent flocking ($\gAA>0$) or nematic order ($\gAA<-0.1$). A disordered homogeneous phase exists within a diamond-shaped region defined by small $|\gAB|$ and $-0.1<\gAA<0$. While this section focuses on the high-density regime to highlight coexistence phases, we note that flocking stripes persist at lower densities and noise levels \cite{lardet2026}. Phase diagrams in the $(D_r,\gAB)$ and $(\rho,\gAB)$ planes are explored in subsequent sections.
Although the phases here are only reported for a fixed system size of $N=2\times10^4$, we performed a finite size analysis in \cite{lardet2026}, confirming that all phases persist up to much larger system sizes, with no noticeable shift in the phase boundaries.

We have tested the robustness of the observed phases to ensure that the patterns are true attractors rather than long-lived transients. First, long numerical simulations confirm that the coexistence phases observed persist over extremely long timescales. We also checked that each phase behavior is obtained from a range of different initial conditions including random (used throughout this study) and purely aligned initial conditions as well as initial conditions in which initial particle orientations are drawn from the von Mises distribution, which is the ordered solution to the Fokker-Planck equation. Finally, we initialized systems using the steady-state solutions taken from across nearby phase boundaries and confirmed that these systems converged to the expected phase behavior. For instance, simulations initialized from a mixed nematic stripe configuration but evolved with parameters corresponding to the antiparallel flocking stripes regime eventually reorganize into the expected anti-flocking pattern.

\begin{figure*}
	\includegraphics[width=\linewidth]{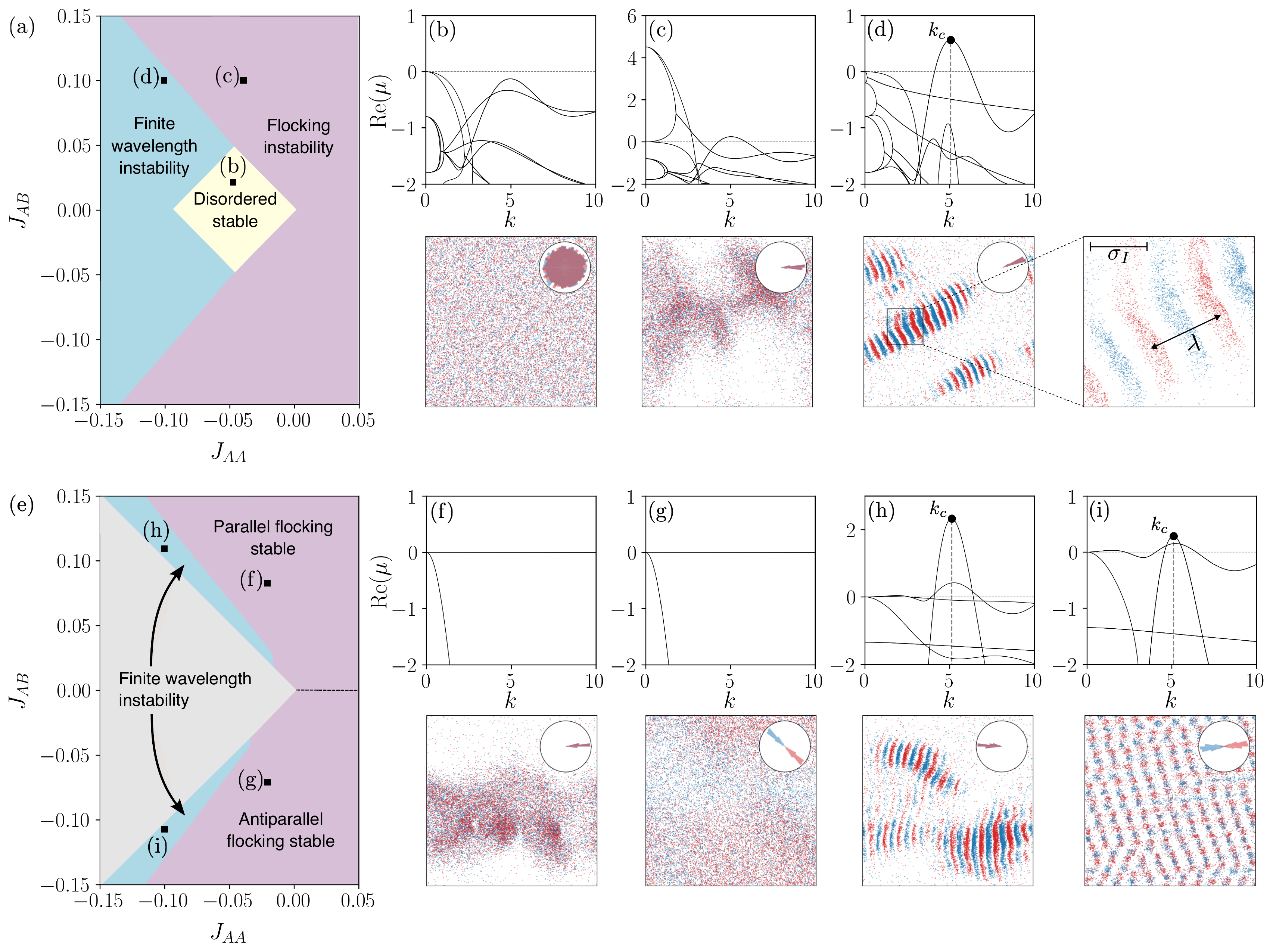}
	\caption{Linear stability analysis results. (a) Phase diagram of linear stability around the disordered homogeneous solution. Yellow indicates the disordered state is stable [$\Re(\mu(k))\leq 0 \ \forall k$]; purple indicates a flocking instability [$\max\{\Re(\mu(k))\}>0$ at $k_{c}=0$]; blue is a finite wavelength instability [$\max\{\Re(\mu(k))\}>0$ at $k_{c}>0$].
	(b)--(d) Growth rates from the disordered state linear stability analysis alongside snapshots from numerical simulations with the same parameters. (b) Disordered stable ($\gAA=-0.05$, $\gAB=0.02$). (c) Flocking instability ($\gAA=-0.04$, $\gAB=0.10$). (d) Finite wavelength instability ($\gAA=-0.10$, $\gAB=0.10$) with $k_c\approx5.1$ indicated with a vertical gray dashed line. There is also a zoomed in snapshot showing the stripe wavelength $\lambda\approx1.23\sigma_I$ in the particle simulations. 
	(e) Phase diagram of linear stability around the ordered homogeneous solution (both parallel and perpendicular to direction of order). Purple indicates the parallel or antiparallel flocking state is stable [$\Re(\mu(k))\leq0 \ \forall k$]; blue is a short-wavelength instability [$\max \Re(\mu(k))>0$ at $k_{c}>2$]; gray indicates no uniform ordered solution from the Fokker-Planck equation exists to perturb around.
	(f)--(i) Growth rates from the ordered state perturbation parallel to direction of travel with $\kbf=(k,0)$ alongside snapshots from numerical simulations with the same parameters. (f) Parallel flocking state is stable ($\gAA=-0.02$, $\gAB=0.08$). (g) Antiparallel flocking state is stable ($\gAA=-0.02$, $\gAB=-0.08$). (h) Finite wavelength instability from parallel ordered state ($\gAA=-0.10$, $\gAB=0.11$). (i) Finite wavelength instability from the antiparallel ordered state ($\gAA=-0.10$, $\gAB=-0.11$).  
	Parameters for linear stability analysis: $N_c=50$, $D_r=0.2$, $v_0=1$, $\RI=1$, $\rholsa=100$. The microscopic simulations were performed with $N=2\times 10^4$ particles.}
	\label{fig:phase_diagram_dis_ord}
\end{figure*}

\subsection{Disordered state perturbation}

We perform a linear stability analysis of the homogeneous disordered state by solving for the eigenvalues of the matrix $\mathcal{L}$ given in Eq.\,(\ref{eq:L_matrix_disordered}). Exploiting the rotational invariance of the disordered state, we restrict the perturbation to the direction $\mathbf{k}=(k, 0)$ without loss of generality. Solving the eigenvalue problem defined by Eqs.\,(\ref{eq:eigenvalue_problem}) and (\ref{eq:lin_op_dis}), truncated at $N_c=50$, yields the growth rates $\mu(k)$. We classify the stability of the system based on the real part of the spectrum:
\begin{enumerate}[(i)]
	\item If $\Re(\mu(k))\leq 0$ for all $k$ then the disordered state is stable;
	\item If $\max\Re(\mu(k))>0$ and is found at $k_{c}=0$, then we expect a system-wide flocking (or antiflocking) instability;
	\item If $\max\Re(\mu(k))>0$ and is found at $k_{c}>0$, then we expect a finite wavelength instability.
\end{enumerate}
The resulting stability phase diagram in the $(\gAA, \gAB)$ plane is presented in Fig.~\ref{fig:phase_diagram_dis_ord}(a), alongside spectra and snapshots from particle simulations corresponding to these three regimes [Figs.~\ref{fig:phase_diagram_dis_ord}(b)--(d)].

We identify a diamond-shaped region defined by $\gAA<0$ and small-to-moderate $|\gAB|$ where the disordered homogeneous state is stable ($\Re(\mu(k))\leq 0$) [Fig.~\ref{fig:phase_diagram_dis_ord}(b)]. Outside this region, instabilities drive the system toward order. When $\gAA>0$ or $|\gAB|$ is large, the spatially homogeneous state becomes unstable to long-wavelength perturbations ($k_c=0$), indicating a flocking instability [Fig.~\ref{fig:phase_diagram_dis_ord}(c)]. Note that while this indicates the onset of polar order, distinguishing between parallel and antiparallel alignment requires analysis of the eigenvector structure or the nonlinear dynamics.

In the remaining region of phase space, the growth rate is maximized at a finite wave number $k_c>0$, indicating a finite-wavelength instability [Fig.~\ref{fig:phase_diagram_dis_ord}(d)]. This predicts the emergence of a pattern with a characteristic wavelength $\lambda \propto 1/k_c$, physically analogous to the length scale selection in Turing patterns~\cite{turing1952}, though arising here via a finite-wavelength Hopf bifurcation---a mechanism often referred to as Turing-Hopf bifurcation \cite{armbruster1988,kepper1994}. 
The coexistence patterns observed in simulations originate from a coupling of the density and polarization modes whereby the density mode becomes dominant at a finite wave number with a nonzero imaginary part. This drives oscillatory behavior of the density field, which in turn induce polarization fluctuations through local (anti-)alignment interactions. This feedback between density and polarization amplifies perturbations at the characteristic wavelength selected by the instability, manifesting as traveling waves.
At higher densities, we also observe a crossover in mode dominance, which may play a role in the stabilization of flocking bands (see Appendix~\ref{sec:appendix_m2_density} for details). Across the parameters tested, the instability occurs systematically in the range $k_c\in [4.995, 5.134]$, corresponding to a characteristic lengthscale $\lambda_{\textrm{LSA}} \in [1.224, 1.258]$, which is strikingly consistent with the wavelength of the traveling phase observed in numerical simulations (see Appendix~\ref{sec:appendix_stripe_wavelength}). 
This emergent wavelength is also consistent with our earlier results for $m=2$ in the mean-sine model \cite{lardet2026}. This indicates that the coexistence patterns are robust against the exact form of the alignment averaging, and rather originate from the competition between the inter- and intraspecies alignment rules.
This mechanism contrasts with the classic Turing instability in which stationary patterns arise as a result of differential diffusion between an activator and inhibitor.

\begin{figure*}[t!]
	\includegraphics[width=\linewidth]{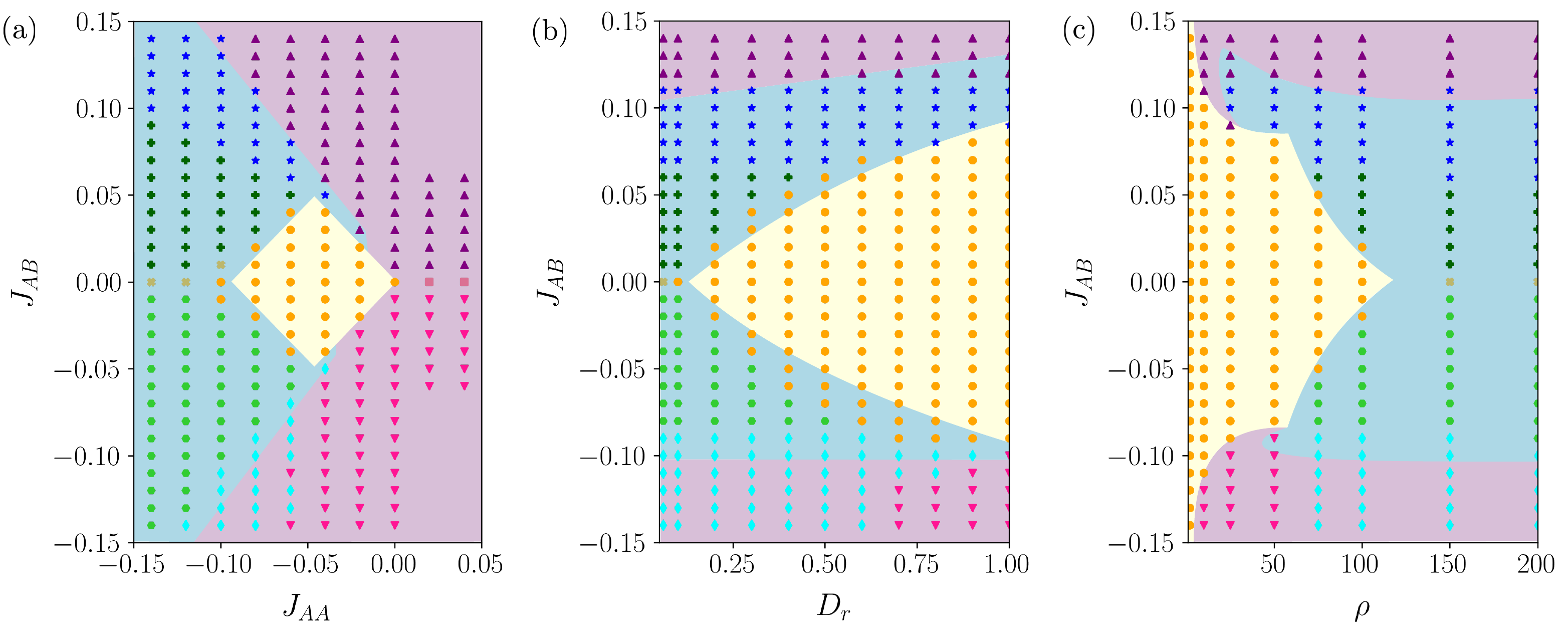}
	\caption{Comparison of linear stability analysis and particle-based simulations. The background colors represent the results of the linear stability analysis combining information from the disordered and ordered state perturbations. Yellow indicates that the homogeneous disordered state is linearly stable to perturbations. Purple indicates that the homogeneous parallel flocking or antiparallel flocking state is stable to perturbations from any direction. Blue indicates a finite wavelength instability in any direction, either from the homogeneous disordered or ordered state.
	The numerical phase classifications are superimposed as scatter points, obtained from particle simulations of the microscopic equations \ref{eq:langevin}. See the caption of Fig.~\ref{fig:phase_diagram_AS_microscopic} for a description of the phase symbols and \cite{lardet2026} for details of the order parameters used for classification. 	The parameters kept constant in each phase diagram are:
	(a) $D_r=0.2$, $\rho=100$; (b) $\gAA=-0.08$, $\rho=100$; (c) $\gAA=-0.08$, $D_r=0.2$. For the linear stability analysis $N_c=50$ and for the particle simulations $N=2\times10^4$.}
	\label{fig:phase_diagram_lsa_overlay}
\end{figure*}

\subsection{Ordered state perturbation} \label{sec:two_sp_ordered_lsa}

We consider a nontrivial polar ordered solution characterized by magnitudes $\Psi^A, \Psi^B > 0$ and orientation angles $\theta_0^A, \theta_0^B$. Defining the complex polarity $\psi^a=\Psi^a e^{i\theta_0^a}$, and aligning $\theta_0^A$ with the $x$-axis without loss of generality, the second species must align either parallel ($\theta_0^B=0$) or antiparallel ($\theta_0^B=\pi$) to the first~\cite{fruchart2021}. Consequently, assuming homogeneous and reciprocal couplings, the system exhibits either parallel flocking ($\psi^A=\psi^B$) or antiparallel flocking ($\psi^A=-\psi^B$).

To determine the existence of these solutions, we examine the mean-field self-consistency relation $x=I_1(ax)/I_0(ax)$, which exhibits a critical point at $a=2$, above which a nontrivial solution $x>0$ appears. For a system with $\rholsa^A=\rholsa^B$, $J_{AA}=J_{BB}$, $J_{AB}=J_{BA}$, a nontrivial ordered solution to the Fokker-Planck equation exists if:
\begin{equation} \label{eq:lsa_ord_boundary}
    \begin{cases}
    \pi\sigma^2 \rholsa^A (J_{AA}+J_{AB}) > 2D_r, & \textrm{parallel flocking, } \\
    \pi\sigma^2 \rholsa^A (J_{AA}-J_{AB}) > 2D_r, & \textrm{antiparallel flocking.}
\end{cases}
\end{equation}
In regimes where both parallel and antiparallel solutions satisfy the existence condition, we select the state with the larger polar order magnitude. This selection rule effectively results in parallel flocking for $\gAB>0$ and antiparallel flocking for $\gAB<0$.

Where an ordered solution exists, we construct the linear stability matrix (Eq.~\ref{eq:lin_op_ordered}), truncated at order $|n|=N_c$, and compute its eigenvalues. Unlike the disordered phase, the ordered state breaks rotational symmetry; thus, one needs to solve this eigenvalue problem for all possible values of $\kbf$ and check for the direction of the largest growing modes. In practice, however, it suffices to check perturbations parallel [$\mathbf{k}=(k,0)$] and perpendicular [$\mathbf{k}=(0,k)$] to the direction of flocking (see Appendix~\ref{sec:appendix_dir_instability} for justification).

The resulting phase diagram is shown in Fig.~\ref{fig:phase_diagram_dis_ord}(e). We classify the stability scenarios as follows:
\begin{enumerate}[(i)]
    \item \textit{Stable Ordered State:} If $\Re(\mu(k))\leq 0$ for all $k$, the homogeneous flocking state (parallel or antiparallel) is stable [Figs.~\ref{fig:phase_diagram_dis_ord}(f)--(g)].
    \item \textit{Finite-Wavelength Instability:} If $\max \Re(\mu(k))>0$ at $k_c>0$, the system displays a finite wavelength instability leading to pattern formation [Figs.~\ref{fig:phase_diagram_dis_ord}(h)--(i)].
\end{enumerate}
Notably, we identify regions of finite-wavelength instability for both $\gAB>0$ and $\gAB<0$ close to the boundary of the region described in Eq.~\ref{eq:lsa_ord_boundary}; these regions were not captured by the stability analysis of the disordered state.

It is important to note that Fig.\,\ref{fig:phase_diagram_dis_ord} primarily highlights \textit{short-wavelength} instabilities. A region of \textit{long-wavelength} instability is found throughout the upper half plane, albeit with very small eigenvalue amplitudes, decaying as we get deeper into the ordered phase (see Appendix~\ref{sec:appendix_long_wavelength}). We also observe a sharp discontinuity in the critical wave number $k_c$ between short-wavelength modes ($k_c\approx 5.1$, $\lambda \approx 1.23$) and long-wavelength modes ($k_c < 1.55$). Given we are concerned with patterns of a short-wavelength with $\lambda\approx 1.23$, we have therefore chosen a minimum cutoff value of $k_c=1.8$. This threshold corresponds to approximately one-quarter of the simulation box size shown in Fig.\,\ref{fig:phase_diagram_dis_ord}, ensuring a clear distinction between microscopic ordering and macroscopic clustering. We also use an instability threshold of $\max\Re(\mu)>10^{-2}$, which is 20 times smaller than the value of $D_r$ used here, ensuring that only genuine instabilities are captured, excluding negligibly small long-wavelength growth rates discussed above. In fact, a weak long-wavelength instability is observed even for effective one-species systems with small coupling values when $\gAA=\gAB$ (see Appendix~\ref{sec:appendix_long_wavelength}).
In regions where no solution to the Fokker-Planck equation exists, the linear stability of the ordered state is naturally undefined.

\subsection{Comparison with numerical simulations}

We now synthesize the results of the linear stability analyses for the disordered and ordered states, comparing them directly to the phases observed in microscopic particle-based simulations. Figure~\ref{fig:phase_diagram_lsa_overlay} presents the combined phase diagrams in the $(\gAA, \gAB)$, $(D_r, \gAB)$, and $(\rho, \gAB)$ planes.

We find that the stability boundary of the disordered state (orange circles) is predicted remarkably well by the linear stability analysis (light yellow region). Similarly, the regimes of parallel and antiparallel flocking (purple and pink symbols) are well captured,  although the emergent microscopic states tend to exhibit significant clustering rather than being spatially homogeneous [see Figs.~\ref{fig:phase_diagram_AS_microscopic}(c) and (f)]. This clustering is a known characteristic of Vicsek models with additive interactions~\cite{chepizhko2021}.

In the remainder of the phase space, the linear theory predicts a finite-wavelength instability with a predicted characteristic lengthscale of $\lambda \approx 1.23$. This value is in excellent agreement with the wavelength measured in simulations, $\lambda=1.222 \pm 0.049$ (see Appendix~\ref{sec:appendix_stripe_wavelength} for further details). This instability region correlates strongly with the coexistence phases found in simulations (blue and green symbols). The transition from homogeneous parallel flocking (purple upward triangle) to the flocking stripes phase (dark-blue star) is in quantitative agreement with the theoretical prediction from the linear stability analysis. Furthermore, both the mixed and demixed nematic lane/stripe phases (green symbols) fall within the predicted finite-wavelength instability region; these states can be further distinguished via a more refined analysis of the linear stability results (see Appendix~\ref{sec:appendix_lsa_mixed}).

Finally, we note that the antiparallel flocking stripes phase (light-blue diamond) is only partially captured by the linear analysis. A portion of this regime is predicted to show stable homogeneous antiparallel flocking. This discrepancy likely arises from nonlinear effects, which may stabilize striped patterns over a uniform ordered state.

\begin{figure*}
	\includegraphics[width=\linewidth]{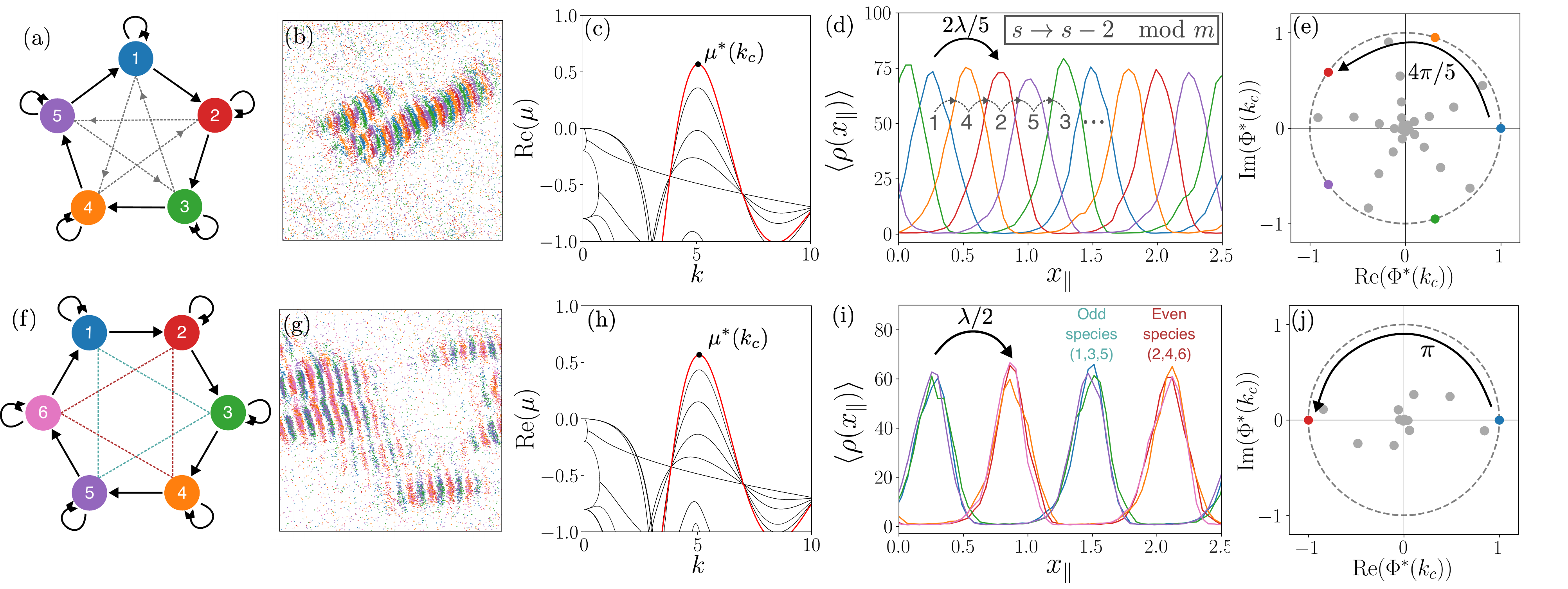}
	\caption{Analysis of the flocking stripes phase with $m=5$ (top) and $m=6$ (bottom) species. (a), (f) Schematics of the cyclic interactions according to Eq.~\ref{eq:cyclic_couplings}.
	The black arrows \alignarrow represent the interspecies alignment and the inverted chevron arrows \antialignarrow represent the intraspecies antialignment. 
	In (a) the gray arrows denote the ordering of the stripes in simulations. In (f) the blue and red triangles denote the grouping of species in the stripes. (b), (g) Snapshots of simulations of the microscopic equations Eq.~\ref{eq:langevin} with parameters $N=2\times10^4$, $\rho=100$, $D_r=0.2$, and $J=0.05m$. (c), (h) Growth rates obtained from linear stability analysis of the disordered state, using the same parameters used in the simulations for (b), (g). Both show a finite wavelength instability with $k_c\approx5$. The growth rate associated with the largest eigenvalue is highlighted in red with the largest eigenvalue given by $\mu^*(k_c)$. (d), (i) Time-averaged density profiles, projected along the direction of motion, are shown for simulations of the microscopic equations with a mean-sine alignment in the flocking stripes phase to explain the stripe ordering ($\rho=100$, $D_r=0.2$, $J=10(m+1)$). As shown in panels (b) and (g), a more clustered version of the flocking stripes phase is observed in the additive-sine model studied in this study. (e), (j) Coefficients of the eigenvector associated with $\mu^*(k_c)$ in the complex plane. The largest magnitude coefficients are colored according to the species they correspond to.}
	\label{fig:odd_even_main}
\end{figure*}

\section{Multi-species alignment cycles} \label{sec:mutli_species_results}
\subsection{Multi-species flocking stripes} \label{sec:multi_stripes}

\begin{figure*}
	\includegraphics[width=\linewidth]{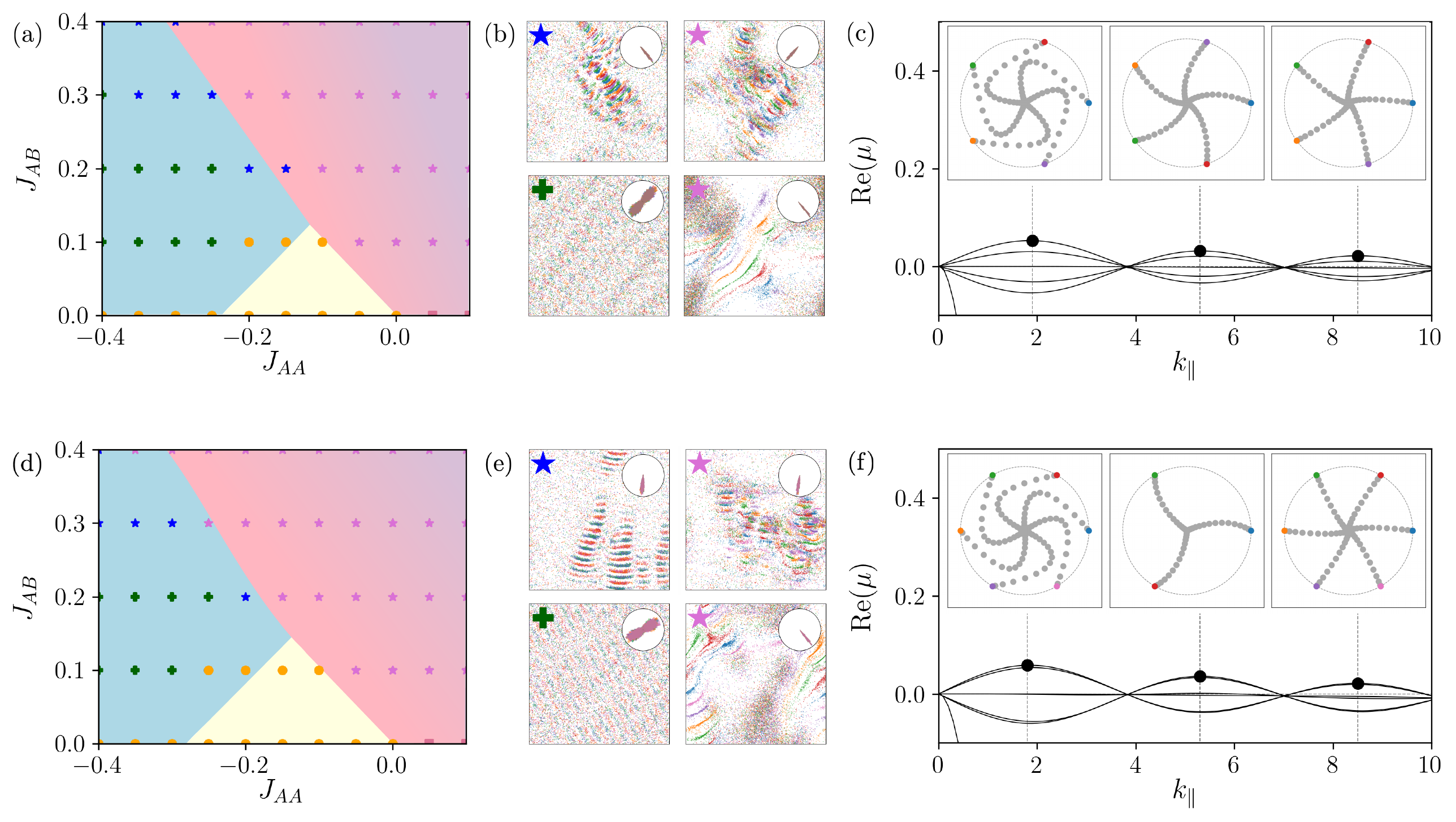}
	\caption{Phase behavior of general multispecies alignment cycles for $m=5$ (a)--(c) and $m=6$ (d)--(f).
  (a), (d) Phase diagrams in $(\gAA,\gAB)$ space with $N=2\times10^4$, $\rho=100$, and $D_r=0.2$. Symbols correspond to numerical simulations---see the caption of Fig.~\ref{fig:phase_diagram_AS_microscopic} for the correspondence between phase behavior and symbols. Pink stars are here representing flocking stripes with different stripe ordering. 
  Background colors correspond to the results of the linear stability analysis, combining information from the disordered and ordered state perturbations as was done in Fig.~\ref{fig:phase_diagram_lsa_overlay} for instance. Yellow indicates that the homogeneous disordered state is linearly stable to perturbations. Blue indicates a finite wavelength instability in any direction, either from the homogeneous disordered or ordered state, with $k_c\approx5.1$. The gradient of pink to purple corresponds to a weak finite wavelength instability from the ordered flocking state, in the $k_\parallel$ direction. 
  (b), (e) Snapshots from simulations (left to right, top to bottom): flocking stripes [blue star, $(\gAA,\gAB)=(-0.35,0.3)$]; flocking stripes with alternative ordering [pink star, $(\gAA,\gAB)=(-0.2,0.3)$]; demixed nematic stripes [dark-green plus, $(\gAA,\gAB)=(-0.35,0.1)$]; long-time flocking stripes with alternative ordering deeper in ordered phase [pink star, $(\gAA,\gAB)=(0.05,0.3)$].
  (c), (f) Growth rates obtained from linear stability analysis of the ordered state, in the $k_\parallel$ direction for the same parameters as in the pink star top right snapshots in (b) and (e), i.e. $(\gAA,\gAB)=(-0.2,0.3)$. The coefficients of the eigenvector associated with each of the mode peaks are visualized in the complex plane. The largest magnitude coefficients are colored according to their species.}
	\label{fig:multipop_phase_diagram}
\end{figure*}

While the existence of flocking stripes in multispecies systems ($m>2$) was established in prior work~\cite{lardet2026}, we now employ analytical tools to elucidate the origin and ordering of these structures.

We consider cyclic alignment interactions where intraspecies couplings are antialigning and each species aligns exclusively with the next in the sequence [see Figs.~\ref{fig:odd_even_main}(a) and (f)]. Defining the index $s \equiv a-1$ for species $a \in \{1, \dots, m\}$, the coupling matrix is given by:
\begin{equation} \label{eq:cyclic_couplings}
    J_{ss'} = \begin{cases}
        -J, & s'=s, \\
        J, & s' =(s+1) \pmod m, \\
        0, & \textrm{otherwise},
    \end{cases}
\end{equation}
where $J>0$ and $s,s' \in \{0, \dots, m-1\}$. In contrast with the binary ($m=2$) case, the couplings are no longer fully reciprocal due to the absence of interactions from species $a$ to $a-1$.

Particle-based simulations confirm that the flocking stripes phase persists for $m>2$ (Figs.~\ref{fig:odd_even_main}(b), (g), and movies S1, S2 in Ref.~\cite{lardet2026b}). While the additive-sine interactions lead to greater clustering than observed in mean-sine models~\cite{lardet2026}, the ordering dynamics remain consistent. Note that for clarity, we utilize the mean-sine model to generate the time-averaged density profiles shown in Figs.~\ref{fig:odd_even_main}(d) and (i). Indeed, it is much easier to produce time-averaged profiles from system-spanning stripes and the ordering and spacing of the species remain consistent for both models.

Much like the two species flocking stripes, linear stability analysis with cyclic alignment couplings for $m>2$ predicts a finite wavelength instability with $k_c\approx5$ [Fig.~\ref{fig:odd_even_main} (c), (h)], matching the emergent stripe wavelength found in simulations. However, the internal structure of these stripes depends critically on the parity of $m$. For {\it odd} values of $m$, the system forms $m$ distinct chasing stripes. Species $s$ chases species $(s-2) \pmod m$, resulting in a spatial separation between sequential species in the interaction cycle of $\Delta x = \lambda(m-1)/(2m)$ [Fig.~\ref{fig:odd_even_main}(d)]. For {\it even} values of $m$, due to the bipartite nature of the interaction graph, species instead group by parity. Odd-numbered and even-numbered species form distinct clusters that flock together, effectively reducing the system to the two-species phenomenology. The separation between sequential species is thus $\Delta x = \lambda/2$.

This parity dependence can be understood intuitively by considering the properties of the circulant matrix. Each species only aligns with the next one in the sequence, generating a directed interaction cycle. Through this cycle, species $s-2$ indirectly aligns with species $s$ via species $s-1$. This continues backwards along the cycle, creating a propagating chain of alignment from $s\rightarrow s-2$. The parity of $m$ then determines how this chain closes. If $m$ is even then there are two distinct chains created through this process, corresponding to the groupings observed in simulations. However, if $m$ is odd then this relation cycles through all species before returning to its starting point, thus producing a single ordered sequence of $m$ chasing stripes.

We rationalize these topological differences by examining the eigenvector $\Phi^*(k_c)$ associated with the most unstable mode, determined by the eigenvalue equation:
\begin{equation} \label{eq:eigvect_kc}
	\mathcal{L} \Phi^*(k_c) = \Phi^*(k_c) \mu^*(k_c).
\end{equation}
Figures~\ref{fig:odd_even_main}(e) and (j) display the coefficients of the eigenvector $\Phi_n^a(k_c)$ in the complex plane for all species $a$ and modes $n$. The eigenvectors are normalized such that the largest coefficient (corresponding to the dominant mode) has unit magnitude (typically $n=0$, or $n=\pm 1$ at higher densities; see Appendix~\ref{sec:appendix_m2_density}). We highlight the coefficients corresponding to the dominant modes. 

We color the coefficients according to the species they correspond to and notice some key differences in structure between odd and even values for $m$. The distribution of these coefficients in the complex plane reveals the mechanism underlying the spatial ordering. For odd $m$, the $m$ largest coefficients correspond to the $m$ roots of unity. The angular separation between species $a$ and $a+1 \pmod m$ is exactly $2\pi (m-1)/(2m)$ [Fig.~\ref{fig:odd_even_main}(e)]. This phase shift in the complex plane maps directly to the spatial shift of $\lambda (m-1)/(2m)$ observed in simulations. Conversely, for even $m$, the coefficients collapse onto the real axis [Fig.~\ref{fig:odd_even_main}(j)]. Sequential species are separated by a phase angle of $\pi$, causing odd-numbered species to cluster at $+1$ and even-numbered species at $-1$. This corresponds to the half-wavelength separation ($\lambda/2$) observed in the bipartite grouping of the even-$m$ simulations.
Precisely the same ordering of stripes was found in the mean-sine model, though with a lower degree of clustering at the same densities, compared to the additive-sine type, as explained in Section~\ref{sec:sim_m2} for the two species case.

While we highlight $m=5$ and $m=6$ here, these parity-dependent behaviors are consistent across $m=2$ to $7$ (see Appendix~\ref{sec:appendix_all_odd_even}). The parity of $m$ also influences the imaginary part of the growth rate $\Im(\mu(k_c))$, as detailed in Appendix~\ref{sec:appendix_multi_density}.

\subsection{General alignment cycles} \label{sec:multi_cyclic_general}
Going further, we consider a more general form of cyclic couplings
\begin{equation} \label{eq:cyclic_couplings_general}
    J_{ss'} = \begin{cases}
        \gAA, & s'=s, \\
        \gAB, & s' =(s+1) \pmod m, \\
        0, & \textrm{otherwise},
    \end{cases}
\end{equation}
where $\gAA$ are the intraspecies couplings and $\gAB$ are the couplings from one species in the cycle to the next.
The corresponding Fokker-Planck equation always has a disordered uniform solution. A parallel ordered solution also exists if $\pi \sigma^2 \rho^A(\gAA+\gAB) > 2D_r$, just as in the two-species case. For even $m$, an antiparallel ordered solution exists with species in the cycle facing in alternating directions (e.g., $\theta_0^a=0$ for $a\in\{A,C,E\}$, and $\theta_0^b=\pi$ for $b\in\{B,D,F\}$ in $m=6$ case). No such antiparallel solution exists for odd $m$. Instead, we expect more complex phases, with greater coupling frustration and oscillatory dynamics. Here, we will only explore the phase space $\gAB\geq0$ for general $m$ species, in order to perturb around the flocking stripes phase observed for $\gAA=-\gAB$. A full exploration of the phases for $m>2$ species is left for future work.

Numerical simulations for $m=5$ and $m=6$ species reproduce the same phases observed for $m=2$ species: disordered, flocking independently, flocking stripes, and demixed nematic stripes (Figs.~\ref{fig:multipop_phase_diagram}(a) and (d)).
However, key differences include an expansion of the disordered region to larger $|\gAA|$ and $|\gAB|$, and a wider region of the flocking stripes phase with strong polar order and demixing, but exhibiting in some cases a different stripe ordering at long times (see pink star snapshots in Fig.~\ref{fig:multipop_phase_diagram}(b) and (e)). 
Rather than following the species ordering $s \rightarrow s-2$, the stripes now arrange in direct species order $s \rightarrow s-1$, as the interspecies alignment dominates over the intraspecies antialignment, favoring a single-step chasing chain.
The stripes are also less distinctive, with a mixture of stripe wavelengths observed, across different realizations. 
We verified that this alternatively ordered flocking stripes phase was also found in simulations of the mean-sine alignment model (Appendix~\ref{sec:appendix_multi_patterns}). Deeper into the demixed nematic stripes phase, we also observed new attractor configurations for the coexistence patterns---see Appendix~\ref{sec:appendix_multi_patterns} for further details.

Linear stability analysis was carried out for both $m=5$ and $m=6$ phase diagrams, based on perturbations from both the uniform disordered state and parallel flocking state, with the results plotted as background colors in Figs.~\ref{fig:multipop_phase_diagram}(a) and (d). As in the two-species case, a stable disordered state is found in pale yellow regions, while finite wavelength instabilities with $k_c\approx5.1$ appear in the light blue regions. In the remainder of the phase space, the ordered state exhibits weakly damped modes along the parallel direction $k_\parallel$, with growth spectra showing multiple peaks at $k\approx\{1.8, 5.3, 8.4, \dots\}$ that decay slowly in amplitude (Figs.~\ref{fig:multipop_phase_diagram}(c) and (f)). This is the same weak long-wavelength instability found earlier in the two-species case (see Sec.~\ref{sec:two_sp_ordered_lsa} and Appendix~\ref{sec:appendix_long_wavelength}); however, a combination of a lower per-species density (since total density was maintained) and the nonreciprocal nature of the cyclic couplings leads to an increase in the magnitude of the growth rates measured (see Appendix~\ref{sec:appendix_multi_mode_dominance}).
Due to the weak growth rates, structures only appear after very long transient times in the homogeneous flocking phase. For long times, they may continue to show both regions of mixed homogeneous flocks and demixed stripes, as can be seen by the snapshots in the lower right panels in Figs.~\ref{fig:multipop_phase_diagram}(b) and (e), which have been run for four times as long as the other realizations.
As the couplings move deeper into the ordered phase, the maximum growth rate $\max \mathrm{Re}(\mu(k))$ approaches zero, reflected in the phase diagrams as a gradual transition from pink to purple background colors. Much deeper in the ordered phase, beyond the parameters shown in the phase diagram, the growth rates become negligible compared to the rotational diffusion rate $D_r$, and the system is effectively stable.
The change of mode dominance from blue to pink regions, and the subsequent decline in mode amplitude are discussed in more detail in Appendix~\ref{sec:appendix_multi_mode_dominance}.

Examining the eigenvectors of the peaks in the growth spectra (Eq.~\ref{eq:eigvect_kc}) reveals the internal ordering within the stripes. Coloring the eigenvector coefficients by species, we observe that for both odd and even $m$, the largest contributions correspond to the $m$ roots of unity, though now with an angular separation of $2\pi/m$ (leftmost insets of Figs.~\ref{fig:multipop_phase_diagram}(c) and (f)). For even $m$, at every second mode peak, the eigenvector structure changes to show an angular separation of $4\pi/m$ between species. 
For all values of $m$, the eigenvector winding direction alternates between successive instability peaks, which is consistent with the oscillatory nature of $J_1(\sigma k)$, which changes sign between peaks. We found all of these eigenvalues have $\Im(\mu)<0$, so the spatial ordering of the stripes is read clockwise round the unit circle (see Appendix~\ref{sec:appendix_multi_stripe_eigenvector}).
Thus, this predicts alternating stripe ordering between peaks of the growth spectra from $s\rightarrow s\pm1$, with the dominant mode favoring $s\rightarrow s-1$, which matches the ordering generally observed in simulations.
The coexistence of multiple weakly unstable modes with similar growth rates thus provides a unified explanation for both the change in stripe ordering and the mixture of wavelengths observed in simulations of the $m>2$ cyclic alignment model for $\gAB \gtrsim |\gAA|$. Higher order symmetry groupings may be possible for larger even $m$; a systematic investigation of this is left for future work.

\section{Conclusion \& Discussion} \label{sec:conclusion}

The study of multiple species with competitive dynamics is a rapidly developing field. In this work, we have derived a kinetic theory for a multispecies generalization of the Vicsek model in which multiple self-propelled particle species interact through distinct (anti)alignment couplings. In doing so, we bridge the gap between microscopic alignment rules and macroscopic pattern formation. By performing a linear stability analysis, we have provided a theoretical foundation for the rich phenomenology recently observed in particle-based simulations~\cite{lardet2026}. Our work shows that multispecies systems can exhibit a wealth of previously unexplored coexistence phases.

We derived a kinetic theory for the model using a Smoluchowski approach and found that this mean-field approach can reliably predict the emergent patterns and instabilities in microscopic particle-based simulations. For binary systems, our analysis successfully delimits the boundaries of the disordered, homogeneous flocking, and coexistence phases. A key result is the identification of a finite-wavelength instability that accurately predicts the emergence of flocking stripes (and laned coexistence phases with intraspecies antialignment). The characteristic length scale derived from the most unstable mode, $\lambda_{\textrm{LSA}} \approx 1.23$, is in excellent quantitative agreement with the stripe periodicity observed in microscopic simulations. This confirms that the formation of these high-density traveling bands is driven by a linear instability mechanism akin to a Turing instability, but mediated here by the nonequilibrium nature of the active driving and alignment couplings via a Hopf bifurcation. 

We extended our theory to cyclic multispecies interactions, in which each species aligns with the next in a directed cycle, while antialigning with members of its own kind. In doing so, we revealed a profound connection between the topology of the interaction network and the spatial structure of the flock. This class of system is directionally frustrated, as no globally consistent order can satisfy all pairwise interactions. As a result, the system resists homogeneous ordered states and instead supports spatially structured dynamics, such as the flocking stripes we observe. We demonstrated that the ``chasing'' behavior of species is encoded in the complex phase of the eigenvectors of the linearized operator. Again, our kinetic theory correctly predicts a characteristic lengthscale, which manifests as the stripe wavelength in simulations of the microscopic equations. 

Depending on the parity of the alignment cycle, we found key differences in the emergent dynamics from both theory and simulations. In odd-length cycles, the spatial separation between consecutive species in the alignment cycle is consistently $\frac{m-1}{2m} \lambda$ where $m$ is the cycle length. In even-length cycles, however, particles form two distinct groups of stripes, with every second species overlapping with one another, resulting in a stripe separation of $\lambda/2$. The spectral analysis offers a clean geometric explanation for the parity effect: for odd interaction cycles, the phases are distributed as roots of unity, mapping directly to a staggered spatial chasing; for even cycles, the phases collapse onto the real axis, forcing a bipartite clustering that reduces the system to an effective two-species phenomenology.
Perturbing the coupling values around these multispecies flocking stripes, we observed similar phases to the binary species system. However, in the regime $\gAB>|\gAA|$, we also uncovered an alternative species ordering of the stripes, which is accompanied by a change in the structure of the eigenvalues and eigenvectors obtained from the linear stability analysis of the ordered state.

We note that the distinction between odd and even cycle dynamics mirrors observations in other active matter systems with cycles of attraction and repulsion \cite{ouazan-reboul2023}, suggesting universality in the underlying mechanisms governing cyclic interactions in active systems. The dynamics of the cyclic model are reminiscent of intransitivity in game theory, such as in the classic game of rock-paper-scissors. This structure has been widely used to study competitive interactions across various socioecological systems \cite{sinervo1996, kirkup2004, wang2014, avelino2018, szolnoki2020}, and can even be a mechanism to sustain coexistence in microbial communities \cite{kerr2002}. This suggests the importance of the role of intransitivity in active matter models, such as ours, not only as a source of stable oscillatory dynamics, but which could also imitate the dynamics of biological processes.

These results highlight the utility of kinetic theory not just for computing phase boundaries, but for decoding the geometry of collective motion in complex networks. 
While the linear stability analysis successfully predicts instability wavelengths and parity-dependent pattern selection, several limitations remain. The chosen coarse-graining method works well for the moderate-to-high densities used here, but is expected to be less useful for low density systems where fluctuations play a greater role.
Moreover, the analysis is restricted to perturbations around stationary uniform base states and therefore cannot describe nonlinear pattern selection, traveling band propagation, or coarsening. Near phase boundaries, weakly nonlinear methods would be needed to characterize the bifurcations more precisely. Perturbations around more complex base states, such as nematic, offset species polar, or spatially inhomogeneous states, are promising directions for further work.

Nevertheless, our kinetic theory framework provides a powerful foundation for understanding emergent phases in multispecies active systems more generally. Future work will leverage this theoretical approach to explore a broader class of systems with complex interaction networks, with the potential of revealing new emergent phases and their self-organizing principles. As experimental realizations of active matter move toward mixtures of synthetic colloids or robotic swarms, our framework provides a general tool for designing interaction rules that target specific self-organized architectures.

\section{Data Availability}
The data that support the findings of this article are openly available \cite{lardet2026b}.

\section{Acknowledgments}
E.L. was funded by a President's Ph.D. Scholarship at Imperial College London. The authors acknowledge computing resources provided by the \href{http://doi.org/10.14469/hpc/2232}{Imperial College Research Computing Service}.

\appendix

\setlength{\tabcolsep}{10pt}
\begin{table*}[t!]
\centering
\begin{tabular}{l|c|c|c|c|c|c|c|c}
& $\Psi$ & $\Psi_A,~\Psi_B$ & $|\delta \theta_{\Psi}|$ & $S$ & $S_A,~S_B$ & $|\delta \theta_{S}|$ & $\overline{d}$ & Periodicity \\ \hline \hline

Parallel flocking 
& $>0$ & $>0$ & $0$ & $>0$ & $>0$ & $0$ & $0$ & No \\ \hline

Antiparallel flocking 
& $0$ & $>0$ & $>0$ & $>0$ & $>0$ & $0$ & $0$ & No \\ \hline

Flocking stripes 
& $>0$ & $>0$ & $0$ & $>0$ & $>0$ & $0$ & $>0$ & Yes \\ \hline

Antiparallel flocking stripes 
& $0$ & $>0$ & $>0$ & $>0$ & $>0$ & $0$ & $>0$ & Yes \\ \hline

Nematic stripes (demixed) 
& $0$ & $0$ & -- & $>0$ & $>0$ & $0$ & $>0$ & Yes \\ \hline

Nematic stripes (mixed) 
& $0$ & $0$ & -- & $>0$ & $>0$ & $0$ & $0$ & Yes \\ \hline

Independent flocking 
& $0$ & $>0$ & $>0$ & $0$ & $0$ & -- & $0$ & No \\ \hline

Independent nematic ordering 
& $0$ & $0$ & -- & $0$ & $>0$ & $>0$ & $>0$ & Yes \\ \hline

Disordered 
& $0$ & $0$ & -- & $0$ & $0$ & -- & $0$ & No \\ \hline

\end{tabular}
\caption{Classification of emergent phases. The definitions of the order parameters are given in Appendix~\ref{sec:appendix_order_params}.}
\label{tab:classification}
\end{table*}

\section{Derivation of $\kappa^a$} 
\label{sec:appendix_kappa}
In this Appendix, we derive the expression for $\kappa^a$ in Eq.~(\ref{eq:kappa}), used in the ordered state perturbation of the two-species system. Starting with the Fokker-Planck equation (\ref{eq:FPE}), we will assume a steady state solution of the form
\begin{equation}
	c_0^a(\theta) = \frac{\rholsa^a}{2\pi} \frac{e^{\kappa^a\cos(\theta-\theta_0^a)}}{I_0(\kappa^a)}.
\end{equation}
Then all temporal and spatial derivatives in Eq.~(\ref{eq:FPE}) will vanish giving us 
\begin{equation} \label{eq:del_A_theta}
	\del_\theta [A(\theta)]=0,
\end{equation}
where 
\begin{align}
	A(\theta) = & \ D_r \del_\theta c_0^a(\theta) \\ & \ - c_0^a(\theta) \pi \sigma^2 \sum_b J_{ab} \int\limits_0^{2\pi} d\theta' c_0^b(\theta') \sin(\theta'-\theta). \nonumber
\end{align}
For this equation to hold for all $\theta$, we require the function inside the derivative to be constant, i.e. $A(\theta)= A_0$. Hence, integrating both sides of $A(\theta)=A_0$ over $\theta$, we get
\begin{align}
	A_0 = -\frac{\sigma^2}{2} \sum_b J_{ab} \int\limits_0^{2\pi}d\theta \int\limits_0^{2\pi} d\theta' c_0^a(\theta) c_0^b(\theta')\sin(\theta'-\theta),
\end{align}
with the diffusion term disappearing due to the periodicity of $c_0^a$. Now, the right-hand side is antisymmetric under a change of variable $\theta \leftrightarrow \theta'$, which forces $A_0= 0$.
Therefore, we have 
\begin{equation} \label{eq:inside_bracket}
	D_r \del_\theta c_0^a(\theta) = c_0^a(\theta) \pi \sigma^2 \sum_b J_{ab} \int\limits_0^{2\pi} d\theta' c_0^b(\theta') \sin(\theta'-\theta).
\end{equation}
For the left-hand side of this equation we have 
\begin{equation}
	D_r \del_\theta c_0^a(\theta) = -D_r \kappa^a \sin(\theta-\theta_0^a) c_0^a(\theta).
\end{equation}
Note that $c_0^a(\theta)$ is strictly positive and so can be canceled on both sides of Eq.\,(\ref{eq:inside_bracket}) giving
\begin{equation}
	-D_r \kappa^a \sin(\theta-\theta_0^a) = \pi \sigma^2 \sum_b J_{ab} \int\limits_0^{2\pi} d\theta' c_0^b(\theta') \sin(\theta'-\theta).
\end{equation}
Let us perform the integral on the right-hand side:
\begin{subequations}
\begin{align}
	K &\equiv \int\limits_0^{2\pi} d\theta' c_0^b(\theta') \sin(\theta'-\theta) \\
	&= \frac{\rholsa^b}{2\pi} \frac{1}{I_0(\kappa^b)} \int\limits_0^{2\pi} d\theta' e^{\kappa^b \cos(\theta'-\theta_0^b)} \sin(\theta'-\theta) \\
	&= \frac{\rholsa^b}{2\pi} \frac{1}{I_0(\kappa^b)} \int\limits_0^{2\pi} d\phi e^{\kappa^b\cos\phi} \sin(\phi - \theta + \theta_0^b),
\end{align}
\end{subequations}
where we have performed a change of variables $\phi=\theta'-\theta_0^b$. Now, using the trigonometric identity $\sin(A+B) = \sin A \cos B + \cos A \sin B$, with $A=\phi$ and $B = \theta_0^b - \theta$, we get
\begin{align}
	K = \ & \frac{\rholsa^b}{2\pi} \frac{1}{I_0(\kappa^b)} \Big[ \cos(\theta_0^b - \theta) \int\limits_0^{2\pi} e^{\kappa^b \cos\phi} \sin\phi  \ d\phi \nonumber \\ 
		&\quad + \sin(\theta_0^b - \theta) \int\limits_0^{2\pi} e^{\kappa^b \cos\phi} \cos\phi  \ d\phi \Big]. 
\end{align}
The first integrand is the product of an even function ($e^{\kappa^b \cos\phi}$) and an odd function ($\sin\phi$), thus its integral over the whole period from $0$ to $2\pi$ is zero. The second integral follows from a known identity. We therefore have
\begin{subequations}
\begin{align}
K &= \frac{\rholsa^b}{2\pi} \frac{1}{I_0(\kappa^b)} 2 \pi I_1(\kappa^b) \sin(\theta_0^b-\theta) \\
&= -\rholsa^b \Psi^b \sin(\theta-\theta_0^b),
\end{align}
\end{subequations}
recalling that $\Psi^b = \frac{I_1(\kappa^b)}{I_0(\kappa^b)}$ is the polar order parameter of species $b$. We thus have the following equality
\begin{equation}
	\sin(\theta-\theta_0^a) \kappa^a = \frac{\pi\sigma^2}{D_r} \sum_b J_{ab} \rholsa^b \Psi^b \sin(\theta-\theta_0^b).
\end{equation}
To remove the dependence on $\theta$, let us project both sides of this equation onto $\sin(\theta-\theta_0^a)$ by integrating over $\theta$. For the left-hand side, we obtain 
\begin{equation}
	\kappa^a \int\limits_0^{2\pi}d\theta \sin^2(\theta-\theta_0^a) = \pi \kappa^a.
\end{equation}
For the right-hand side, we obtain 
\begin{equation}
	\frac{\pi \sigma^2}{D_r} \sum_b J_{ab} \rholsa^b \Psi^b \int\limits_0^{2\pi} d\theta \sin(\theta-\theta_0^a) \sin(\theta-\theta_0^b) = \pi \cos(\theta_0^b - \theta_0^a).
\end{equation}
Therefore, all together we derive the following explicit expression for $\kappa^a$
\begin{equation}
	\kappa^a = \frac{\pi\sigma^2}{D_r} \sum_b J_{ab} \rholsa^b \Psi^b \cos(\theta_0^b - \theta_0^a).
\end{equation}

\section{Order parameters for phase classification} \label{sec:appendix_order_params}

Here we provide a brief overview of the quantities used to classify the phases in Fig.~\ref{fig:phase_diagram_AS_microscopic} (polar order, nematic order, demixing and periodicity). Further details on the calculation of the order parameters and classification of phases can be found in Ref.\,\cite{lardet2026}.

For a set $\Omega$ of particles, the \emph{polar order} and \emph{nematic order} of the system are given by
\begin{align}
	\Psi_{\Omega} &= \left \langle \frac{1}{|\Omega|} \left|\sum_{i \in \Omega} \hat{\mathbf{p}}_i(t) \right| \right \rangle, \\
	S_{\Omega} &= \left \langle \frac{1}{|\Omega|}\left|\sum_{i \in \Omega} \hat{\mathbf{q}}_i(t) \right| \right \rangle,
\end{align}
where $|\Omega|$ is the number of particles in the set $\Omega$ and the angular brackets denote an average over time, once in steady state. Here, $\hat{\mathbf{p}}_i(t)=(\cos\theta_i(t), \sin\theta_i(t))$ is the unit polar orientation vector and $\hat{\mathbf{q}}_i(t)=(\cos2\theta_i(t), \sin2\theta_i(t))$ is the unit nematic orientation vector.
We measure the polar and nematic order within each species $\Psi_A$, $\Psi_B$, $S_A$, $S_B$ as well as the overall order $\Psi$, $S$. We also measure the difference in polar/nematic direction between the two species $\delta\theta_\Psi$ and $\delta\theta_S$.

In order to quantify the local species-species phase separation, we measure the \emph{demixing order parameter}:
\begin{equation}
\overline{d} = \left \langle \frac{1}{N} \sum_{i=1}^N \frac{1}{|\mathcal{D}_i|} \sum_{j\in\mathcal{D}_i} \delta_{s(i),s(j)}-\frac{1}{2}\right \rangle,
\end{equation} 
where $\mathcal{D}_i=\{j:|\rbf_i-\rbf_j|\leq \sigma_I/2 \}$ and $\delta_{s(i)s(j)}$ is a Kronecker delta.
This quantity measures the average ratio of same species particles to total number of neighbors within a radius of $\sigma_I/2$. This distance was chosen such that $\mathcal{D}_i$ is almost only comprised of particles of the same species in the striped coexistence phases. 
Note that $1/2$ is subtracted in the order parameter calculation so that $\overline{d}>0$ describes a demixed binary system and $\overline{d}=0$ describes a completely mixed system.

Finally, we also measure any periodicity displayed by the system to more accurately distinguish micropatterned states and macroscopic flocks (i.e. between the liquid antiparallel flocking and antiparallel flocking stripes phases as the demixing parameter remains nonzero for both). 
A system is labeled as periodic if we observe a large peak in the vicinity of $k = 1/\lambda\approx 0.8$ in the power spectrum $|\delta\hat{\rho}(k)|^2$ of the density fluctuations parallel to the direction of order. The classification of the nine phases using these order parameters is summarized in Table~\ref{tab:classification}.

\begin{figure}
	\includegraphics[width=\linewidth]{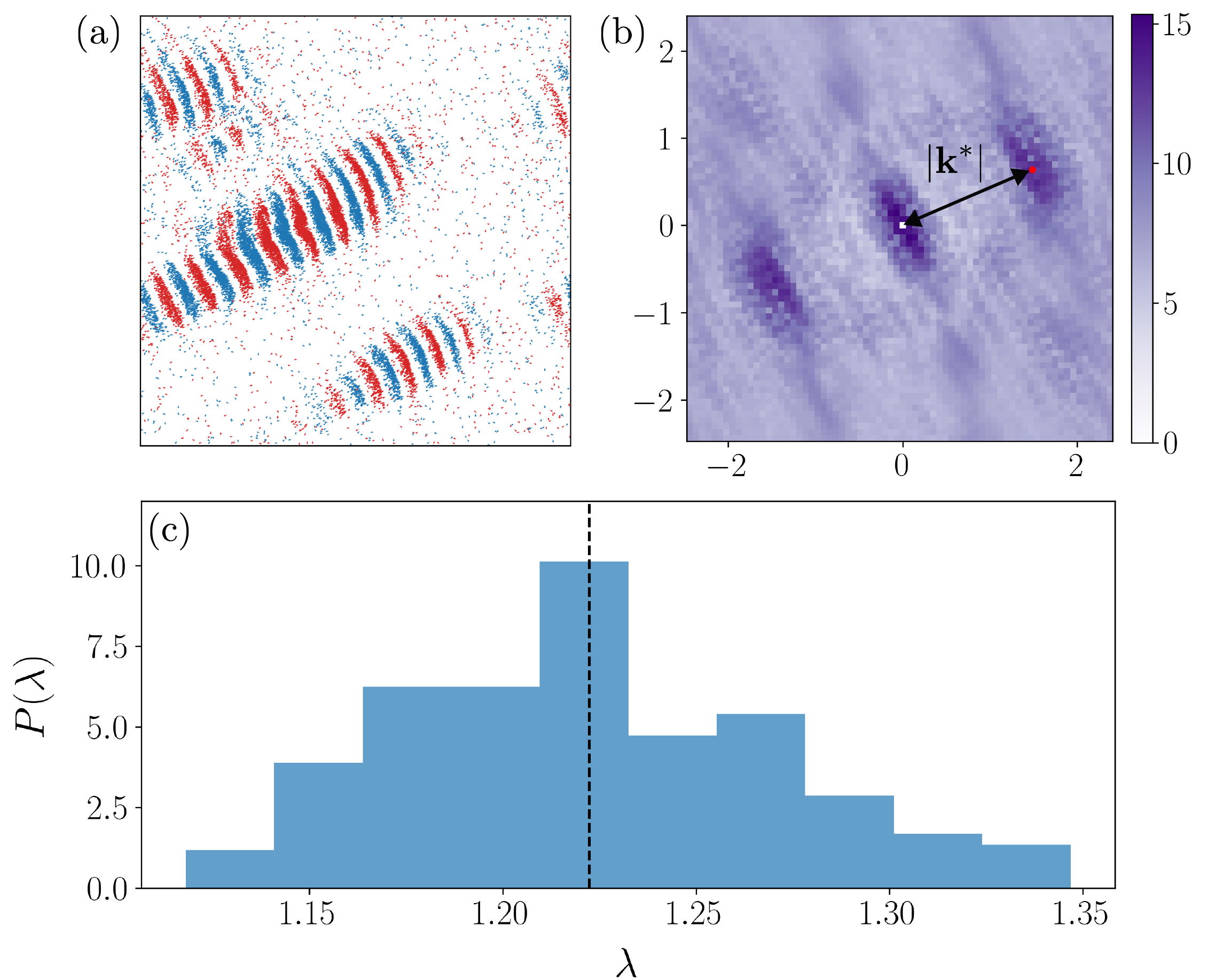}
	\caption{Wavelength of flocking stripes in particle-based simulations of the microscopic equations. (a) Snapshot of a system with flocking stripes and (b) its structure factor. The location of the peak $\kbf^*$ is highlighted in red. (c) Histogram of the same-species stripe wavelength $\lambda=2/|\kbf^*|$ for systems classified as parallel flocking stripes in Fig.~\ref{fig:phase_diagram_AS_microscopic}. The mean wavelength of $\lambda=1.22$ is shown as a vertical dashed black line.}
	\label{fig:wavelength_AS}
\end{figure}

\section{Stripe wavelength in simulations} \label{sec:appendix_stripe_wavelength}

Here, we examine the structure factor to measure the stripe wavelength from the particle-based simulations of the microscopic equations [Eqs.\,(\ref{eq:langevin_r})--(\ref{eq:langevin_theta})]. We define the structure factor $\mathcal{S}(\kbf)$ as the Fourier transform of the density fluctuation correlation function
\begin{equation}
	\mathcal{S}(\mathbf{k}) = \langle |\delta\hat{\rho}(\mathbf{k}, t)|^2 \rangle_t.
\end{equation}
Here, $\delta\hat{\rho}(\mathbf{k}, t)$ is the Fourier transform of the density fluctuations $\delta\rho(\rbf, t)=\rho(\rbf,t) - \langle \rho(\rbf,t)\rangle_{\rbf}$, where $\rho(\rbf,t)$ is the local density measured in square grids of length $\ell=0.2$, centered on $\rbf$, from a coarse-grained snapshot of the system at time $t$.
We then look for a peak of the structure factor in Fourier space $\kbf^*$, which would correspond to a characteristic lengthscale in the system [Fig.~\ref{fig:wavelength_AS}(b)]. We exclude the region $|\kbf^*|<0.5$ as this would correspond to lengths on the scale of the macroscopic clusters rather than the stripe wavelength---see Fig.~\ref{fig:wavelength_AS}(a)--(b).
The average stripe separation in real space will be given by $1/|\kbf^*|$. Defining the stripe wavelength to be the distance between density peaks of the same species, we have $\lambda = 2/|\kbf^*|$ as the stripes are of alternating species. We measured the stripe wavelength for all the systems classified as parallel flocking stripes in Fig.~\ref{fig:phase_diagram_AS_microscopic}, with 10 realizations for each, and show a histogram of their values in Fig.~\ref{fig:wavelength_AS}(c).  
The mean stripe wavelength is measured to be $\lambda=1.222 \pm 0.049$. This is in quantitative agreement with the wavelength found for the mean-sine model simulations of $\lambda\approx1.23$ \cite{lardet2026}.

\section{Mode dominance} \label{sec:appendix_lsa_density}

\subsection{Two-species systems} \label{sec:appendix_m2_density}

\begin{figure*}[p]
	\includegraphics[width=\linewidth]{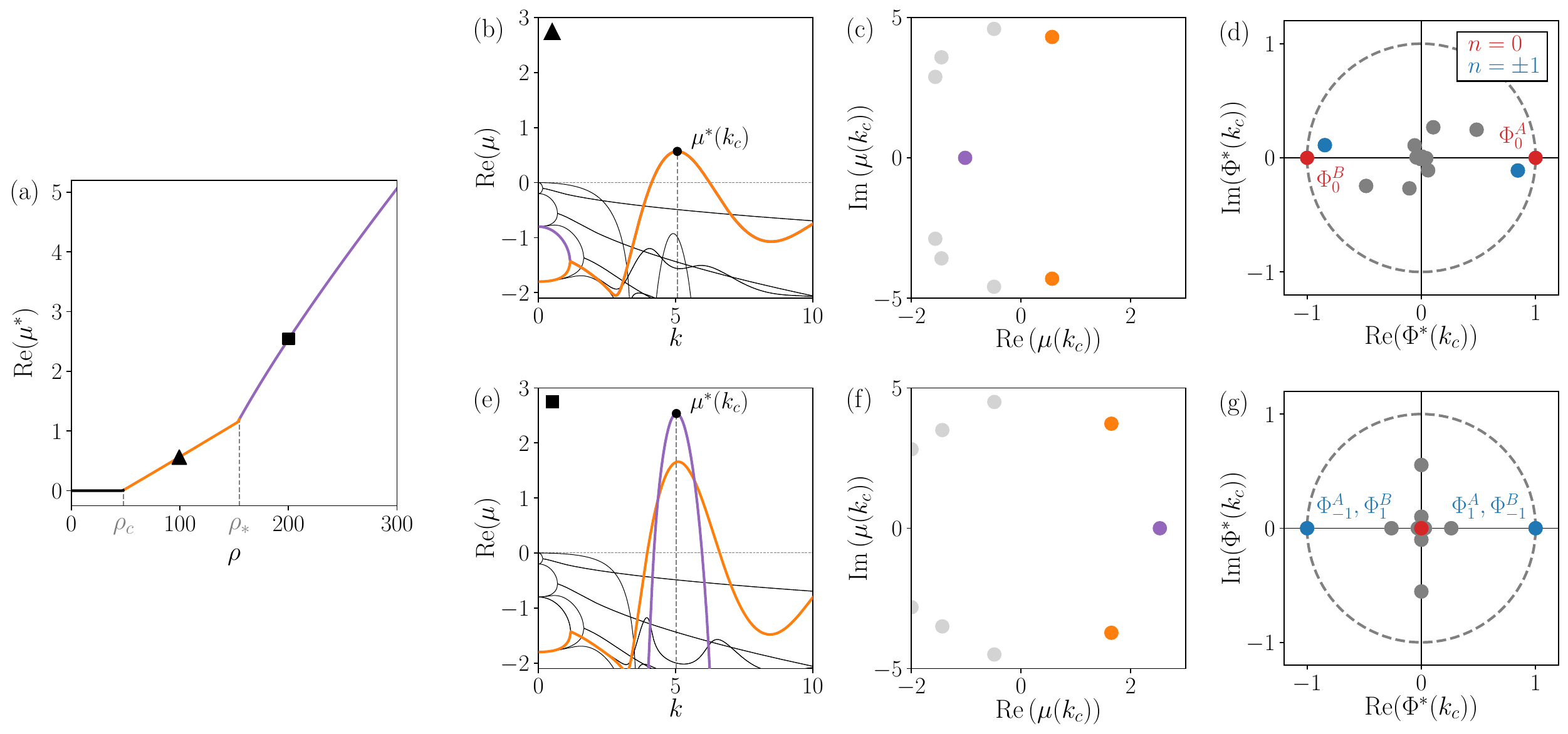}
	\caption{Mode dominance of the finite wavelength instability. (a) Eigenvalue with the largest real part as a function of density. (b), (e) Growth rates at the triangle ($\rho=100$) and square ($\rho=200$) marker indicated in (a). At lower density, the orange growth rate dominates and at higher density the purple growth rate dominates. (c), (f) Eigenvalues at $k_c$ in the complex plane. At lower density, the dominant eigenvalue is a pair of complex conjugates (orange points in (c)). At higher density, the dominant eigenvalue is real with no imaginary part (purple point in (f)). (d), (g) Coefficients of the eigenvector associated with $\mu^*(k_c)$ in the complex plane. The coefficients associated with mode $n=0$ are colored in red, while those for modes $n=\pm1$ are colored in blue. The higher order modes are in gray. At lower density (d), the $n=0$ mode dominates. At higher density (g), the $n=\pm1$ modes dominate. Other parameters were $D_r=0.2$, $\gAA=-0.10$, $\gAB=0.10$ and $N_c=50$.}
	\label{fig:m2_increase_rho}
\end{figure*}

To elucidate the physical mechanism driving the finite-wavelength instability, we analyze the eigenvector structure associated with the largest growth rate. These results are summarized in Fig.~\ref{fig:m2_increase_rho}.

At low densities ($\rho < \rho_c$), the spectrum of the disordered state lies entirely in the left half-plane, indicating stability. As $\rho$ exceeds a critical threshold $\rho_c$, an eigenvalue with positive real part emerges signaling a finite-wavelength instability at $k_c \approx 5$ [Fig.~\ref{fig:m2_increase_rho}(b)]. The leading eigenvalue forms a complex conjugate pair [Fig.~\ref{fig:m2_increase_rho}(c)], and the corresponding eigenvector is dominated by the $n=0$ mode, i.e. the density mode [Fig.~\ref{fig:m2_increase_rho}(d)].
As the density is increased further, beyond a crossover value $\rho_*$, a different mode becomes dominant. While the critical wave number remains approximately constant ($k_c \approx 5$), the growth rate becomes purely real [Figs.~\ref{fig:m2_increase_rho}(e)--(f)]. Concurrently, the eigenvector structure shifts, becoming dominated by the $n=\pm 1$ modes, i.e. the polarity modes [Fig.~\ref{fig:m2_increase_rho}(g)]. Formally, these results indicate that the system undergoes a Hopf bifurcation of type $I_o$ at $\rho_c$~\cite{cross1993}. This is followed by a mode dominance crossover at $\rho^*$ from the density mode to the polarity modes.

The critical density $\rho_c \approx 50$ aligns well with the transition from disorder to flocking stripes observed in particle-based simulations. Regarding the crossover at $\rho_*$, we hypothesize that systems with $\rho_c<\rho<\rho^*$ exhibit an oscillatory density instability (due to the density mode dominating and $\Im(\mu^*(k_c))\neq0$), triggering the local demixing of species. These fluctuations may locally push the density above $\rho_*$, thereby triggering the polarity-dominated mode ($\Im(\mu^*(k_c))=0$) and stabilizing the flocking bands. We note that we did not observe distinct macroscopic signatures distinguishing the regimes below and above $\rho_*$ in the steady-state simulations.

\begin{figure*}[p]
	\includegraphics[width=\linewidth]{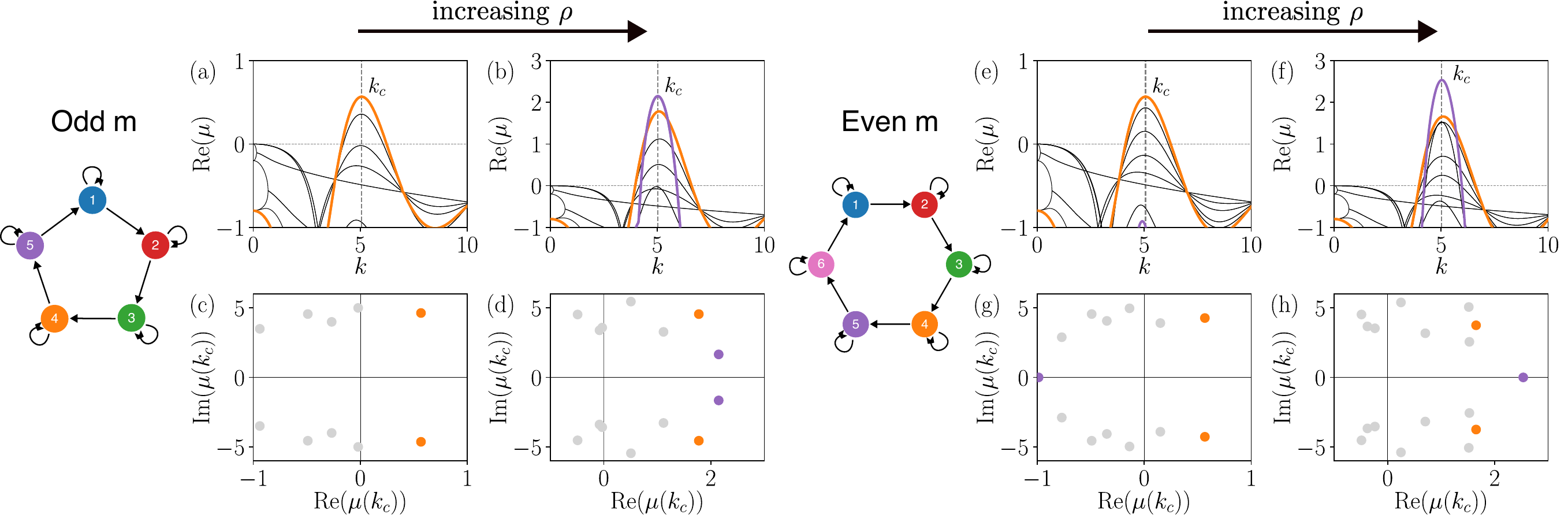}
	\caption{Dominant eigenvalue in odd vs even cycles. On the left side are results for $m=5$ and on the right for $m=6$. Growth rates at $\rho=100$ [(a), (e)] and $\rho=200$ [(b), (f)]. In the panels below each of these are the eigenvalues in the complex plane at $k=k_c$. At lower density for both $m=5$ and $m=6$, the eigenvalues with the largest real part are a pair of complex conjugates [orange points in (c), (g)]. At higher density, the eigenvalues with the largest real part are complex conjugate for $m=5$ [purple points in (d)], but collapse onto the real line for $m=6$ [purple point in (h)]. Other parameters were $D_r=0.2$, $J=0.05m$ and $N_c=50$.}
	\label{fig:odd_even_increase_rho}
\end{figure*}
\clearpage

\subsection{Multi-species systems} \label{sec:appendix_multi_density}
In multispecies systems, examining the structure of the eigenvectors corresponding to the largest growth rate reveals important differences between odd and even $m$ [see Fig.~\ref{fig:odd_even_increase_rho}]. We restrict our analysis to the unstable regime $\rho > \rho_c$, where the system exhibits a finite-wavelength instability [Figs.~\ref{fig:odd_even_increase_rho}(a) and (e)].

In the intermediate density regime $\rho_c < \rho < \rho_*$, the linear stability landscape is qualitatively similar across all species counts: the leading eigenvalue $\mu_{\max}$ forms a complex conjugate pair for both odd and even $m$ [Figs.~\ref{fig:odd_even_increase_rho}(c) and (g)]. Beyond $\rho_*$, we observe a mode dominance crossover analogous to the two-species case [Figs.~\ref{fig:odd_even_increase_rho}(b) and (f)]. However, the spectral character of this new dominant mode depends strictly on the parity of $m$. For odd $m$, the growth rate $\mu(k_c)$ remains a complex conjugate pair. In contrast, for even $m$, the eigenvalue becomes purely real [Figs.~\ref{fig:odd_even_increase_rho}(d) and (h)].

\section{Direction of instability} \label{sec:appendix_dir_instability}
\begin{figure}
	\includegraphics[width=\linewidth]{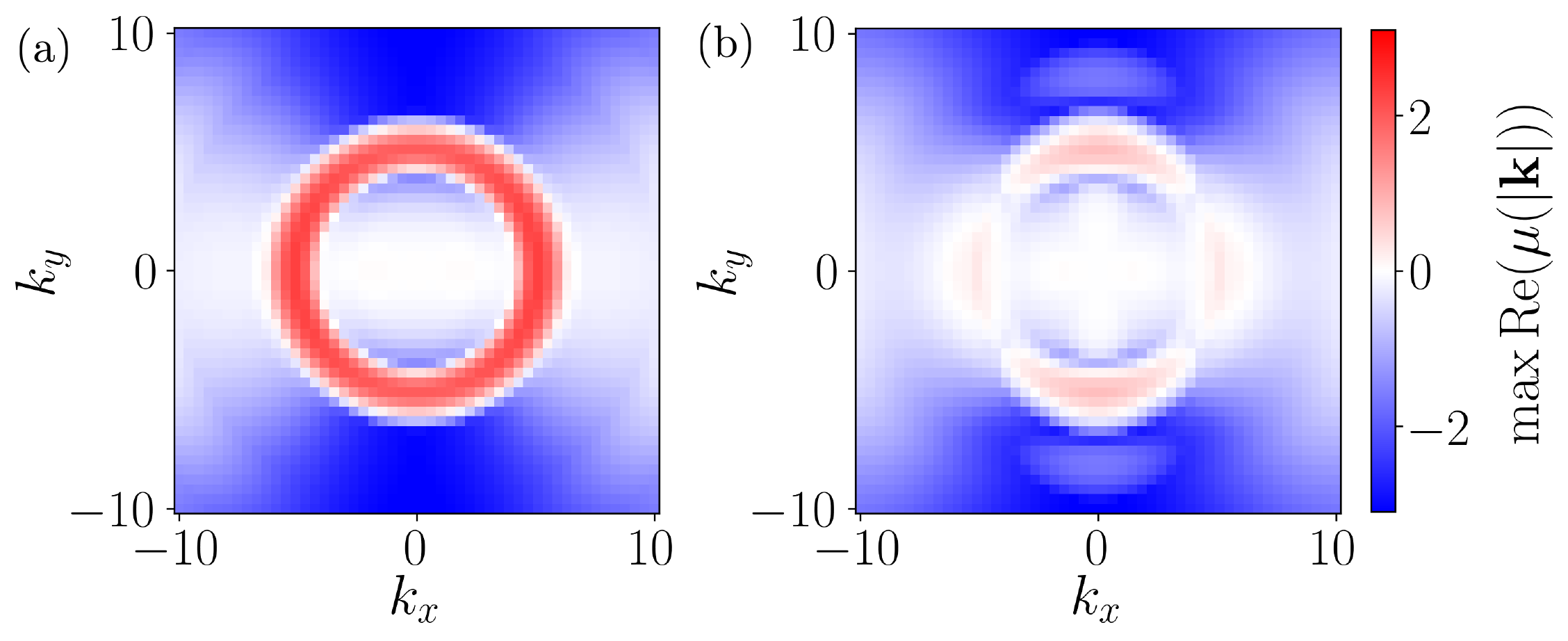}
	\caption{Eigenvalue with maximum real part over all $(k_x, k_y)$ values for linear stability from the ordered homogeneous state (parallel flocking if $\gAA>0$ or antiparallel flocking if $\gAB<0$).
	Coupling parameters were (a) $\gAA=-0.10$, $\gAB=0.11$ and (b) $\gAA=-0.10$, $\gAB=-0.11$. Other parameters were $D_r=0.2$, $\rho=100$ and $N_c=50$.}
	\label{fig:k_dir}
\end{figure}

In Fig.\,\ref{fig:k_dir}, we show $\max\big\{\Re(\mu(|\kbf|))\big\}$ in all directions $(k_x,k_y)$ for the linear stability from the homogeneous ordered state. We see that either the fastest growing mode grows in all direction uniformly (a), or in the direction parallel or perpendicular to the direction of order (b), depending on the parameters. It is therefore sufficient to check the linear stability in the directions $(k_x,0)$ and $(0,k_y)$, which allows us to greatly save on computation time.

We tried starting particle-based simulations with an initial configuration following a von Mises distribution [Eq.~(\ref{eq:von_mises})] and found that the initial instability, indeed, appears to either start in the direction parallel or perpendicular to the direction of polar/nematic order (see movies S3 and S4 in Ref.~\cite{lardet2026b}, which were initialized from von Mises distributions with parallel and antiparallel flocking, respectively).

\begin{figure}
	\includegraphics[width=0.8\linewidth]{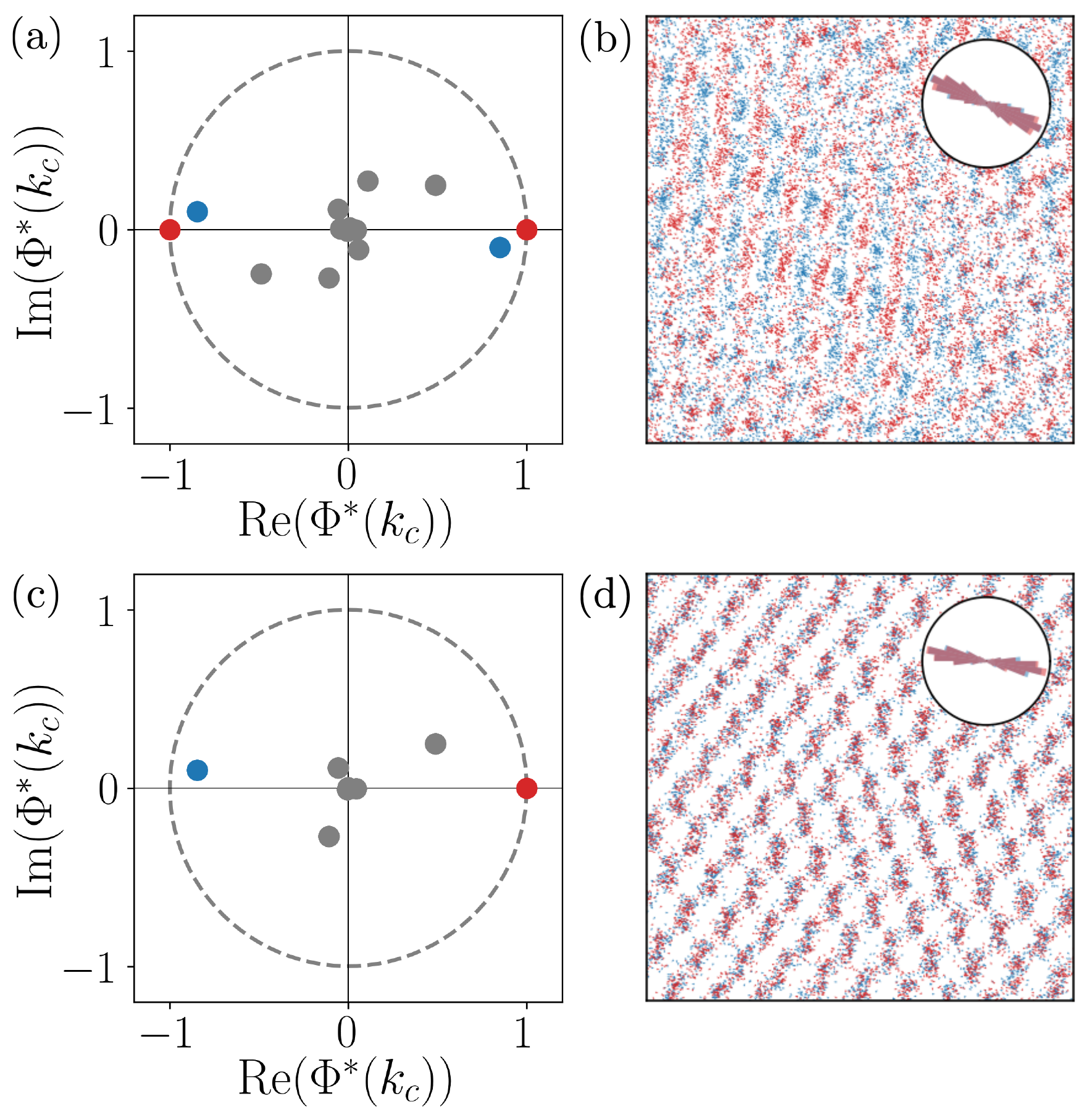}
	\caption{Mixed and demixed phases linear stability comparison. (a), (c) Coefficients of the eigenvector associated with $\mu^*(k_c)$ in the complex plane for (a) $\gAA=-0.14$, $\gAB=0.05$ and (c) $\gAA=-0.14$, $\gAB=-0.05$. The coefficients associated with mode $n=0$ are colored in red and those for modes $n=\pm1$ are colored in blue. The higher order modes are gray. The other parameters are $D_r=0.2$, $\rho=100$ and $N_c=50$. To the right are snapshots with the same parameters from simulations with $N=2\times 10^4$ particles for (b) $\gAA=-0.14$, $\gAB=0.05$ displaying nematic stripes with species that are demixed and (d) $\gAA=-0.14$, $\gAB=-0.05$ displaying nematic stripes with species that are mixed.}
	\label{fig:mixed_eigvects_compare}
\end{figure}

\section{Weak long-wavelength instability} \label{sec:appendix_long_wavelength}

\begin{figure*}
	\includegraphics[width=\linewidth]{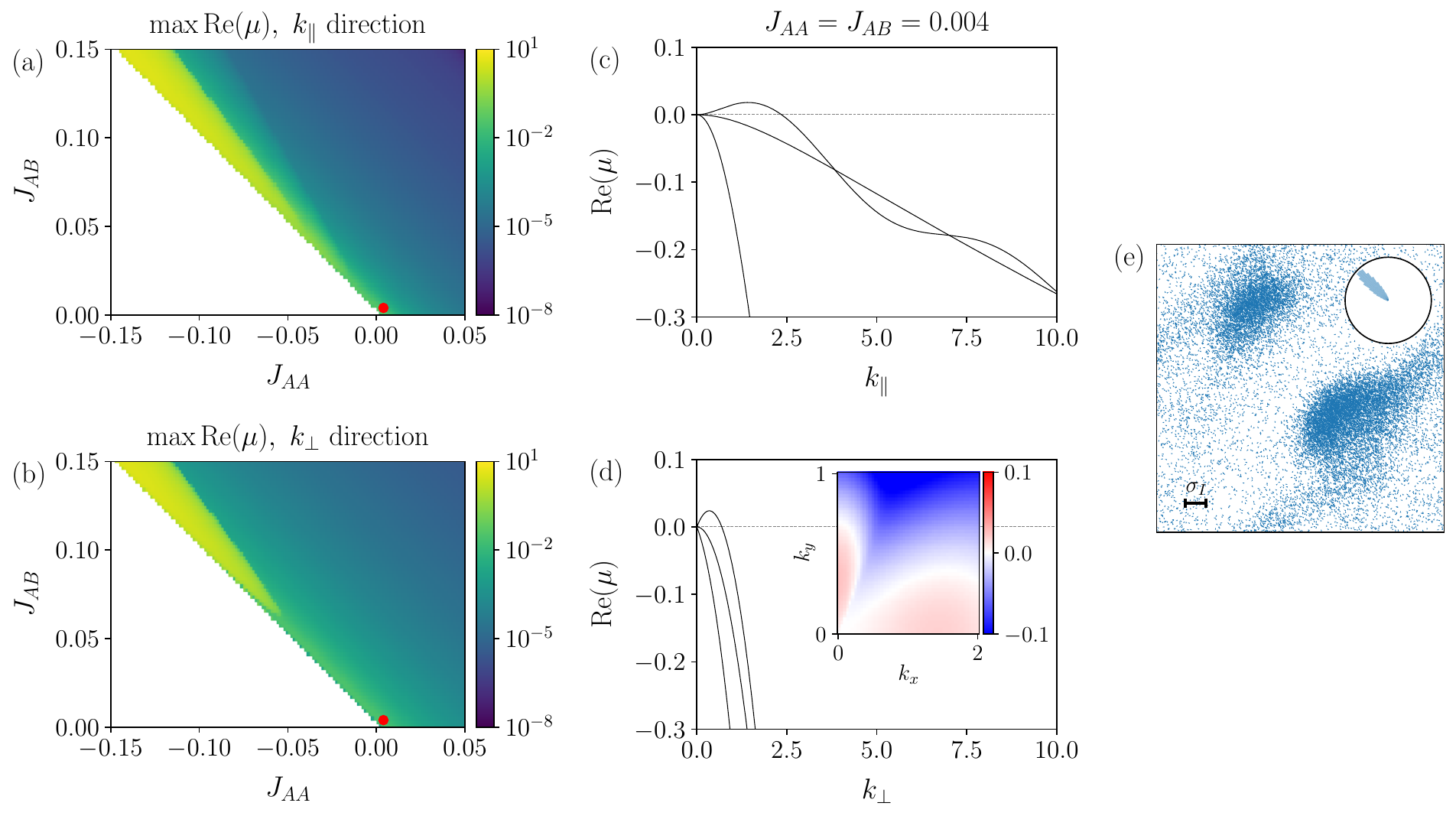}
	\caption{Long wavelength instability in linear stability analysis of the ordered state. (a), (b) Phase diagrams in the ($\gAA, \gAB$) plane of the eigenvalue with maximum real part, with a logarithm scale, in both the $k_\parallel$ (a) and $k_\perp$ (b) directions. (c), (d) Growth rates for the point circled in red in panels (a) and (b) with $\gAA=\gAB=0.004$ (effectively one species) showing a long-wavelength instability of $\Re(\mu(k_\parallel))=0.018$ at $k_\parallel=1.42$ and $\Re(\mu(k_\perp))=0.024$ at $k_\perp=0.31$. Other parameters were $D_r=0.2$, $\rho=100$ and $N_c=50$. The inset of (d) shows $\max\Re(\mu(\kbf))$ in $(k_x,k_y)$ space. (e) Snapshot of a particle-based simulation of the additive-sine model for $m=1$ with $N=2\times10^5$ and the same other parameters as used to generate (c) and (d).
	}
	\label{fig:long_wavelength}
\end{figure*}

In the main text, we were mainly concerned with the short-wavelength perturbation with $k\approx 5.1$ corresponding to the emergent wavelength of stripes and lanes in the simulations. However, we also identify that the ordered flocking state has a weak long wavelength instability in both the parallel and perpendicular directions in the upper half plane for $m=2$ species---see Fig.\,\ref{fig:long_wavelength}.
The flocking striped state has a large amplitude finite wavelength instability at $k\approx 5$. Moving to larger $\gAA$ from here, we find that the ordered state perturbation has a weak amplitude finite wavelength instability in both directions, but at longer wavelengths $0<k<2$. The peak of this instability decays quickly as we go deeper into the ordered phase $\gAA, \gAB>0$. We also verified that $(k_\parallel,0)$ and $(0,k_\perp)$ were the two most unstable directions in the inset to Fig.\,\ref{fig:long_wavelength}(d), similar to what we found in Appendix~\ref{sec:appendix_dir_instability}.

Interestingly, a long-wavelength instability is observed even for effective one-species systems. As an example, we choose a system close to the order-disorder transition, such that its instability amplitude is not too small in magnitude, as shown in Fig.~\ref{fig:long_wavelength}(c)--(d), and structures may therefore be visible in a reasonable simulation time. Running a realization of this system for long times, we indeed begin to observe long-wavelength clustering in this one-species system with the same parameters [Fig.~\ref{fig:long_wavelength}(e)]. 

Given the extremely weak growth rate, we expect that when $\max\Re(\mu)\ll D_r$, these instabilities will be negligible on the timescale of rotational diffusion. For the purposes of classifying the phase diagrams in the main text, we use an instability threshold of $\max\Re(\mu)>10^{-2}$, which is 20 times smaller than the value of $D_r$ used in most of the simulations. In Fig.~\ref{fig:phase_diagram_dis_ord}, this results in essentially the whole weak long-wavelength instability region being classified as parallel flocking stable (purple).

This weak amplitude instability is discussed again in more detail for $m>2$ species in Appendix~\ref{sec:appendix_multi_mode_dominance}.

\section{Difference between mixed and demixed stripes} \label{sec:appendix_lsa_mixed}

Within the region of finite-wavelength instability of the disordered state in the two-species system, we identify a clear distinction between mixed and demixed striped phases based on the structure of the unstable eigenmodes.
We examine the eigenvector coefficients of the most unstable mode in the complex plane. For parameters corresponding to demixed states in the particle-based simulations, such as flocking stripes or demixed nematic stripes [Fig.~\ref{fig:mixed_eigvects_compare}(b)], the dominant coefficients for the two species are separated by an angular distance of $\pi$, lying on the real axis at $\pm 1$ [Fig.~\ref{fig:mixed_eigvects_compare}(a)]. This corresponds to a spatial separation of $\lambda/2$ between stripes of alternating species.
In contrast, for parameter values displaying the mixed nematic stripe phase in simulations [Fig.~\ref{fig:mixed_eigvects_compare}(d)], the dominant coefficients coincide in phase at $\Phi^*(k_c)=1$ for both species. In this case, the two species are in phase within each stripe, leading to a stripe spacing of $\lambda$ and mixing between species.

\section{Odd vs even flocking stripes} \label{sec:appendix_all_odd_even}

\begin{figure*}[p]
	\includegraphics[width=\linewidth]{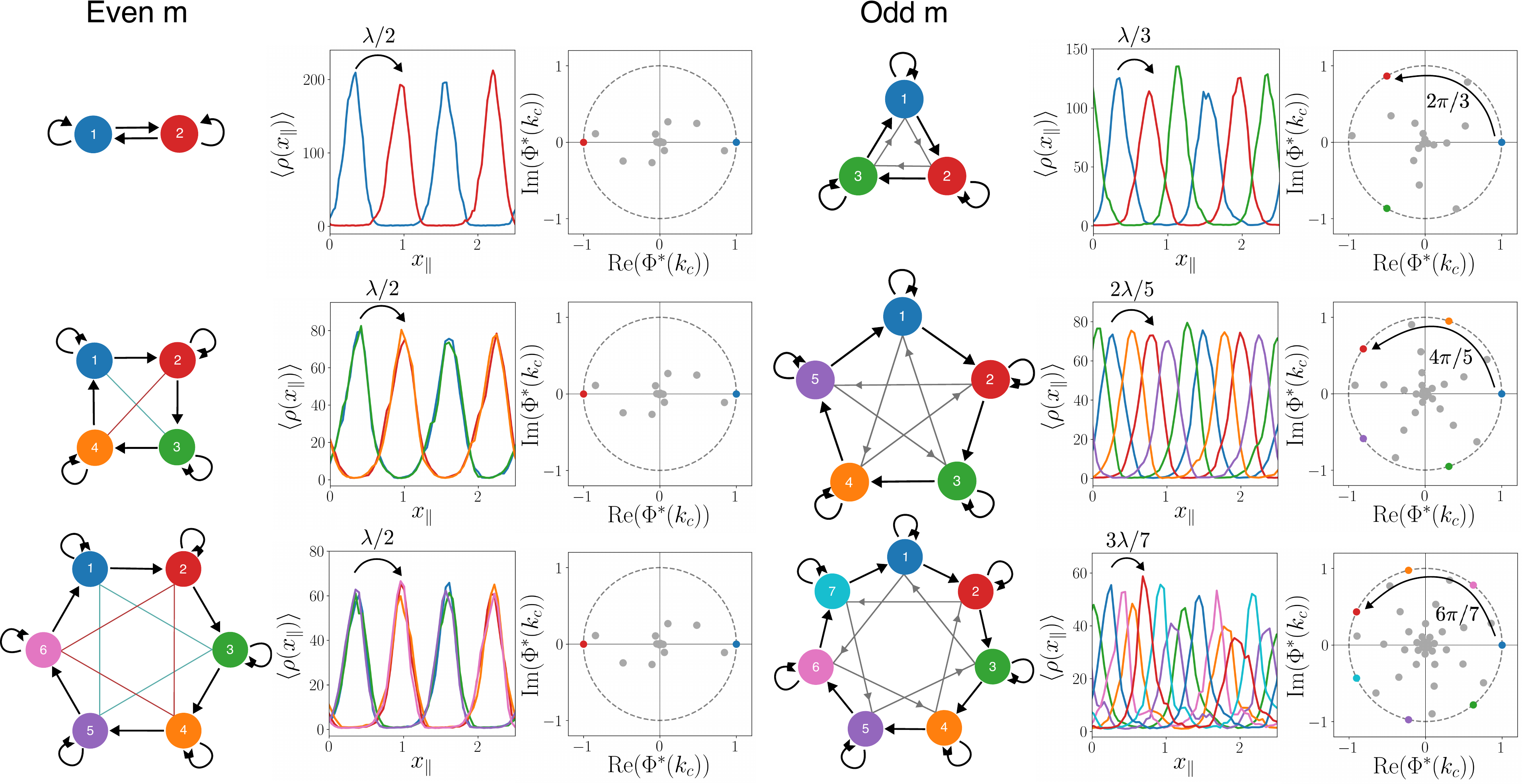}
	\caption{Comparison of odd and even numbers of species with cyclic alignment interactions. For each $m=2$ to $m=7$ three panels are shown. On the left, a schematic of the cyclic interactions is shown. 
	The black arrows \alignarrow represent the interspecies alignment and the inverted chevron arrows \antialignarrow represent the intraspecies antialignment. 
	For even $m$, the blue and red lines show the grouping of species. For odd $m$, the gray lines show the ordering of the stripes. In the center, the time-averaged density profiles, projected along the direction of travel, are shown for simulations of the microscopic equations with a mean-sine alignment in the flocking stripes phase. The distance from one species to the one it is aligning with is given as a fraction of $\lambda$, the stripe wavelength of the same species. Simulation parameters were: $N=5m\times10^4$, $\rho=100$, $D_r=0.2$, $J=10(m+1)$.
	On the right, from linear stability analysis of the disordered state in the additive-sine interaction model, the coefficients of the eigenvector corresponding to the largest eigenvalue $\Phi^*_n(k_c)$ are plotted in the complex plane. The coefficients of mode $n$ with the largest magnitude are colored according to the species number they correspond to, in accordance with the colors used in the interaction schematics. The linear stability analysis parameters were: $N_c=50$, $\rho=100$, $D_r=0.2$, $J=0.05m$.}
	\label{fig:odd_even_all}
\end{figure*}

In the main text, we reported differences in the stripe configuration between $m=5$ and $m=6$ species with alignment cycles, citing their differences being due to the parity of the cycle length. In Fig.~\ref{fig:odd_even_all}, we perform the linear stability analysis from the disordered state on all systems with species numbers from $m=2$ to $m=7$, showing clear differences in the stripe ordering and eigenvector structure depending on the parity of $m$.

\clearpage

\section{Further multispecies coexistence patterns} \label{sec:appendix_other_multi}

\subsection{Microscopic simulations} \label{sec:appendix_multi_patterns}
\begin{figure}
	\includegraphics[width=\linewidth]{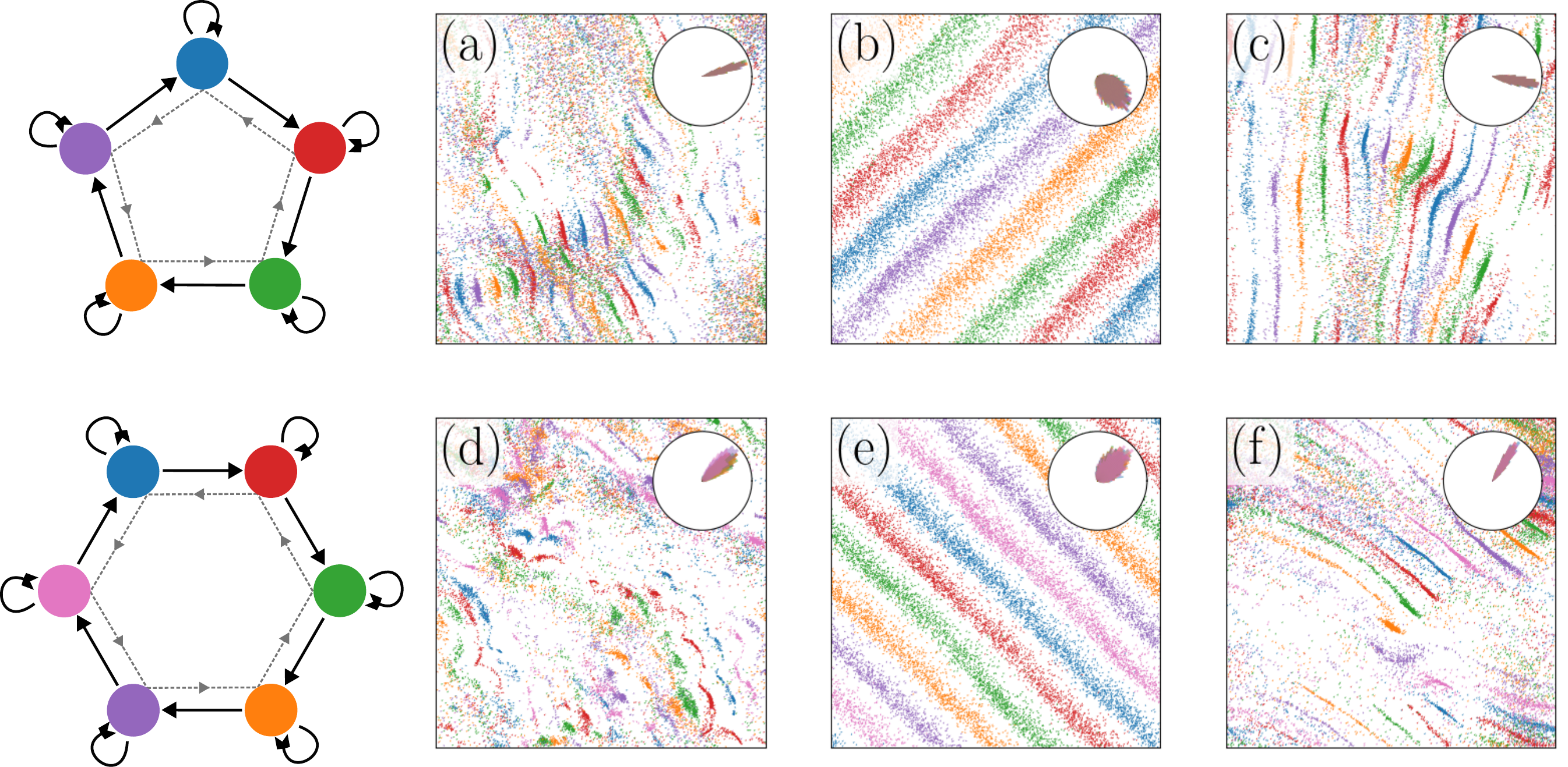}
	\caption{Snapshots of multispecies flocking stripes with alternative ordering in the mean-sine alignment model. (a)--(c) have $m=5$ species and (d)--(f) have $m=6$ species. The schematic in each row shows the cyclic alignment rules in black, and the emergent ordering of the stripes in gray dashed arrows, going in the opposite direction to the cycle. Coupling parameters are: (a) $\gAA=-30$, $\gAB=100$; (b) $\gAA=0$, $\gAB=20$; (c) $\gAA=0$, $\gAB=40$; (d) $\gAA=-40$, $\gAB=100$; (e) $\gAA=0$, $\gAB=20$; (f) $\gAA=0$, $\gAB=60$. Other simulation parameters were $N=2\times10^4$, $D_r=0.2$ and $\rho=100$.}
	\label{fig:multi_stripes_MS}
\end{figure}

We verified that the alternative ordering of the multispecies flocking stripes when $\gAB\gtrsim |\gAA|$ is also observed in the mean-sine type alignment model. Some example snapshots are shown in Fig.~\ref{fig:multi_stripes_MS}, clearly displaying the stripe ordering of $s\rightarrow s-1$ in both odd and even cases. A variety of emergent stripe wavelengths are observed, despite all simulations being carried out with the same box size. Interestingly, demixing can even occur when $\gAA\geq0$ (i.e. no longer intraspecies antialignment). This is not completely unexpected since the $m>2$ cyclically aligning systems are in general nonreciprocal, which has been shown recently to be a generic mechanism for demixing in active alignment models \cite{myin2026}.

\begin{figure}
	\includegraphics[width=\linewidth]{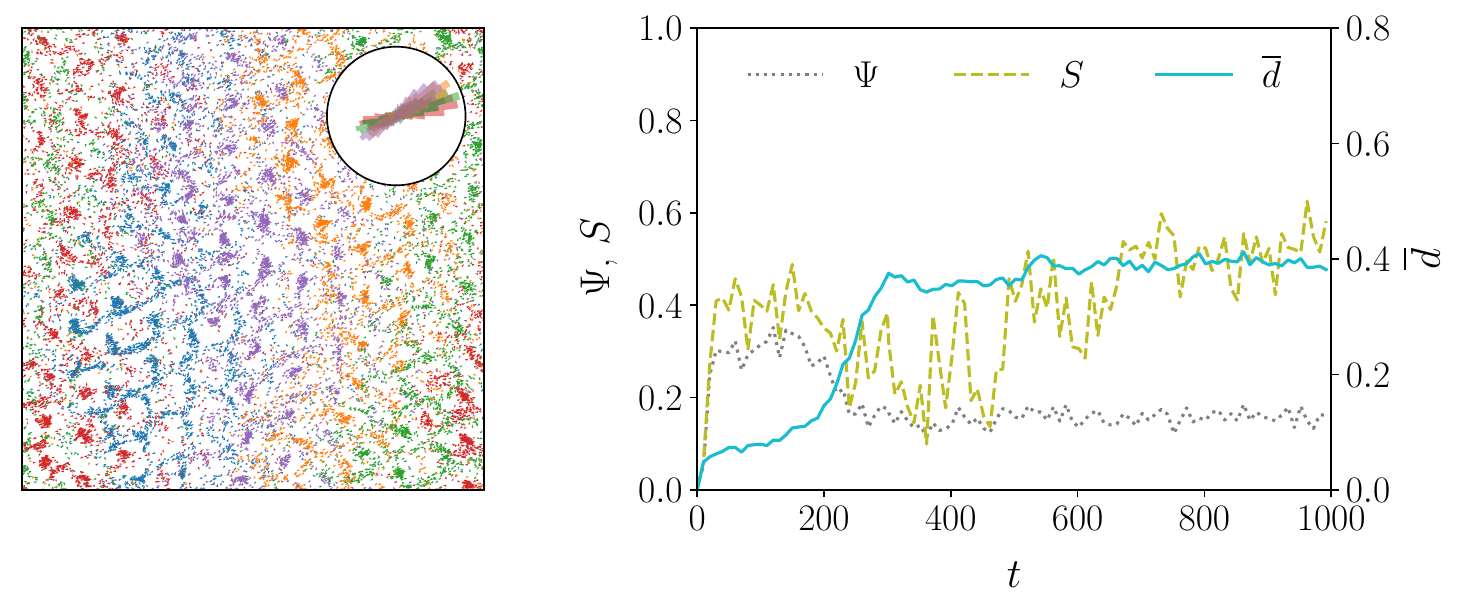}
	\caption{Additional demixed nematic coexistence pattern at long times for $m>2$. A snapshot is shown on the left, while the evolution of its order parameters over time is shown in the right panel (polar order $\Psi$, nematic order $S$ and demixing $\bar{d}$ order parameter). Note that for $m>2$, we subtract $1/m$ within the order parameter calculation such that $\overline{d}=0$ is still the threshold for a demixed system. As such the maximum value is now $\overline{d}=1-1/m$. Parameters used were $m=5$, $N=2\times10^4$, $D_r=0.2$, $\rho=100$, $\gAA=-0.4$ and $\gAB=0.3$.}
	\label{fig:demix_new_pattern}
\end{figure}

In $m>2$ systems, we observed a new type of coexistence pattern for strong intraspecies antialignment and interspecies alignment (top left of phase diagrams in Fig.~\ref{fig:multipop_phase_diagram}). Compared to our original demixed nematic stripes, we observe, at long times, that these systems tend to demix even further and form larger scale single species regions (Fig.~\ref{fig:demix_new_pattern}).
We verified that this pattern is an attractor, by initializing some simulations from the final configuration of nearby phases in the phase diagram to check that this configuration was still reached.

\subsection{Mode dominance from the ordered perturbation} \label{sec:appendix_multi_mode_dominance}
\begin{figure*}
	\includegraphics[width=\linewidth]{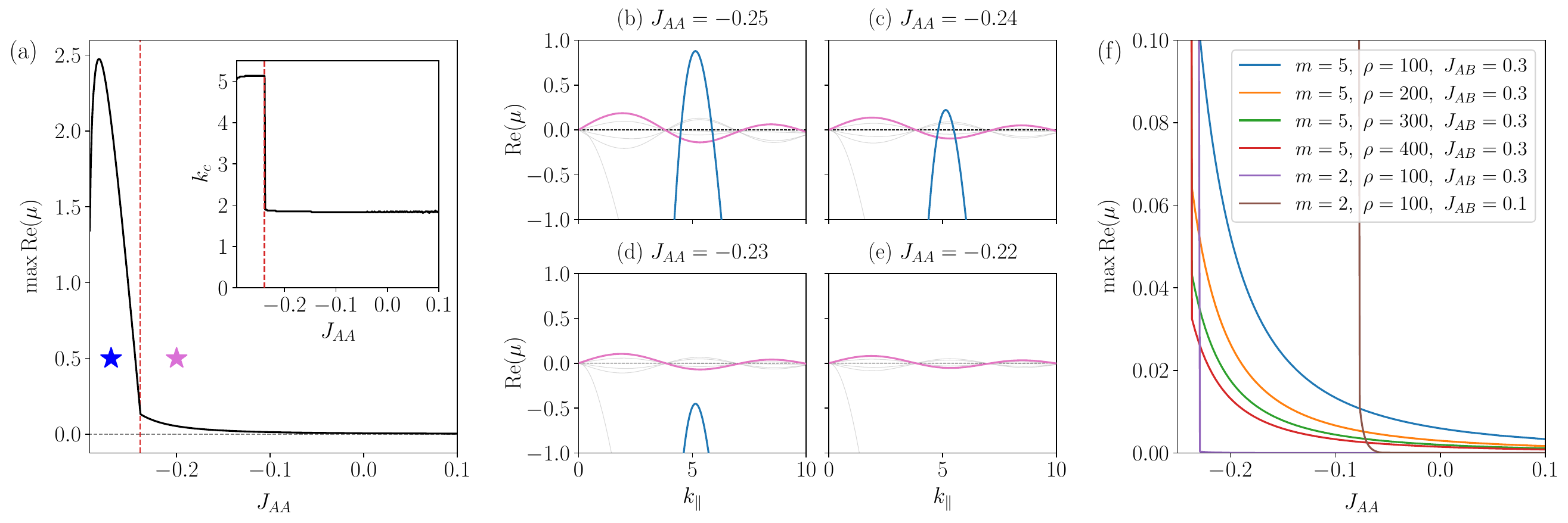}
	\caption{Eigenvalues from the linear stability around the ordered state, looking in the $k_\parallel$ direction with $N_c=100$. (a) Most unstable eigenvalue $\max\mathrm{Re}(\mu)$ across a slice of the phase diagram in Fig.~\ref{fig:multipop_phase_diagram} at $\gAB=0.3$. An ordered solution only exists for $\gAA\gtrapprox-0.29$. The inset shows the corresponding value of $k_c$ at which $\max\mathrm{Re}(\mu)$ occurs. There is a sharp kink in the maximum eigenvalue at $\gAA=-0.239$, at which point the value of $k_c$ has a sharp discontinuity (plotted as a red dashed vertical line in both plots). Growth spectra across this transition are shown in (b)--(d), with the blue and pink modes tracked across each value of $\gAA$.
  (f) Most unstable eigenvalue $\max\mathrm{Re}(\mu)$, varying total system density $\rho$ and species number $m$, zoomed in on small amplitudes. Growth rates decrease as density is increased. They decay even further in the ordered phase for $m=2$.}
	\label{fig:multi_stripes_eigs_track}
\end{figure*}

In Sec.~\ref{sec:multi_cyclic_general}, we observed a transition in eigenvalue behavior in the linear stability analysis around the ordered state, which was also accompanied by a change in stripe ordering in the particle-based simulations.
Here, we investigate further this change in mode dominance across a slice of the phase diagram in Fig.~\ref{fig:multipop_phase_diagram}(a) at $\gAB=0.3$ for $m=5$. In Fig.~\ref{fig:multi_stripes_eigs_track}, we observe the sharply peaked mode at $k_c\approx5$ decreases in amplitude as $\gAA$ increases, shown in blue in panels (b)--(d). For $\gAA\gtrapprox-0.239$, the dominant mode switches to the weakly damped oscillatory mode, colored in pink, which has its maximum at $k_c\approx 2$. The discontinuity across this transition is clearly observed in Fig.~\ref{fig:multipop_phase_diagram}(a) in both the maximum real part of the eigenvalue spectrum, and its corresponding wave number location. This transition corresponds well with the change in stripe behavior we observed in simulations, which we color as blue star and pink stars.
As we increase $\gAA$ further, the weakly damped oscillatory mode continues to decrease in magnitude. 
In simulations, we thus begin to observe systems that take much longer to demix.

Finally, in Fig.~\ref{fig:multi_stripes_eigs_track}(f), we examine the magnitude of $\max\Re(\mu)$ as density and the number of species are varied. We clearly see that the $m=5$ growth rates are much larger compared to the $m=2$ case in the ordered phase. For $m=5$, the growth rates decrease as density is increased. However, the $m=5$ systems still have larger growth rates than $m=2$, even at comparable or higher per-species densities. We hypothesize that this is due to nontrivial effects of the nonreciprocal coupling network, leading to more unstable states. This comparison supports our findings of an enhanced region of longer wavelength stripe instabilities in the $m>2$ systems shown as a pink to purple shading in Fig.~\ref{fig:multipop_phase_diagram}. In comparison, for $m=2$, the growth rates decrease sharply to much smaller magnitudes at the transition from flocking stripes to homogeneous parallel flocking states, hence the solid purple color in Figs.~\ref{fig:phase_diagram_dis_ord} and \ref{fig:phase_diagram_lsa_overlay}.

\subsection{Stripe ordering from the complex eigenvector} \label{sec:appendix_multi_stripe_eigenvector}

From Eqs.\,(\ref{eq:perturbation_fourier}) and (\ref{eq:perturbation_fourier_ordered}), the density perturbation for species $a$ at wave number $k$, in the parallel direction, takes the form
\begin{equation}
    \delta\rho^a(x,t) \propto e^{\Re(\mu)t}\cos\bigl(\Im(\mu)\,t + kx + \phi_a\bigr),
\end{equation}
where $\phi_a = \arg(\Phi_0^a)$ is the argument of the density component of the eigenvector. The density peak of species $a$ sits where the cosine is maximized:
\begin{equation}
    x_a(t) = \frac{-\Im(\mu)\,t - \phi_a}{k}.
\end{equation}
The stripe velocity is $\dot{x}_a = -\Im(\mu)/k$, so for $k>0$ the direction of travel is determined by the sign of $\Im(\mu)$. At any fixed time, the species ahead in the direction of travel are those with smaller $\phi_a$ when $\Im(\mu)<0$, and those with larger $\phi_a$ when $\Im(\mu)>0$. On the unit circle, decreasing argument happens in the clockwise direction and increasing argument in the anticlockwise direction, so:
\begin{itemize}
  \item for $\Im(\mu)<0$, eigenvector coefficients should be read in the \textit{clockwise} direction;
  \item for $\Im(\mu)>0$, eigenvector coefficients should be read in the \textit{anticlockwise} direction.
\end{itemize}
For the instability around the disordered state for the flocking stripes in Fig.~\ref{fig:odd_even_main}, the dominant density mode corresponds to a pair of complex conjugate eigenvalues (see Figs.~\ref{fig:m2_increase_rho} and \ref{fig:odd_even_increase_rho}). We always took the eigenvalue with $\Im(\mu)>0$, so the eigenvector coefficients are read anticlockwise, giving $s \to s-2$ stripe ordering.

\begin{figure}
	\includegraphics[width=\linewidth]{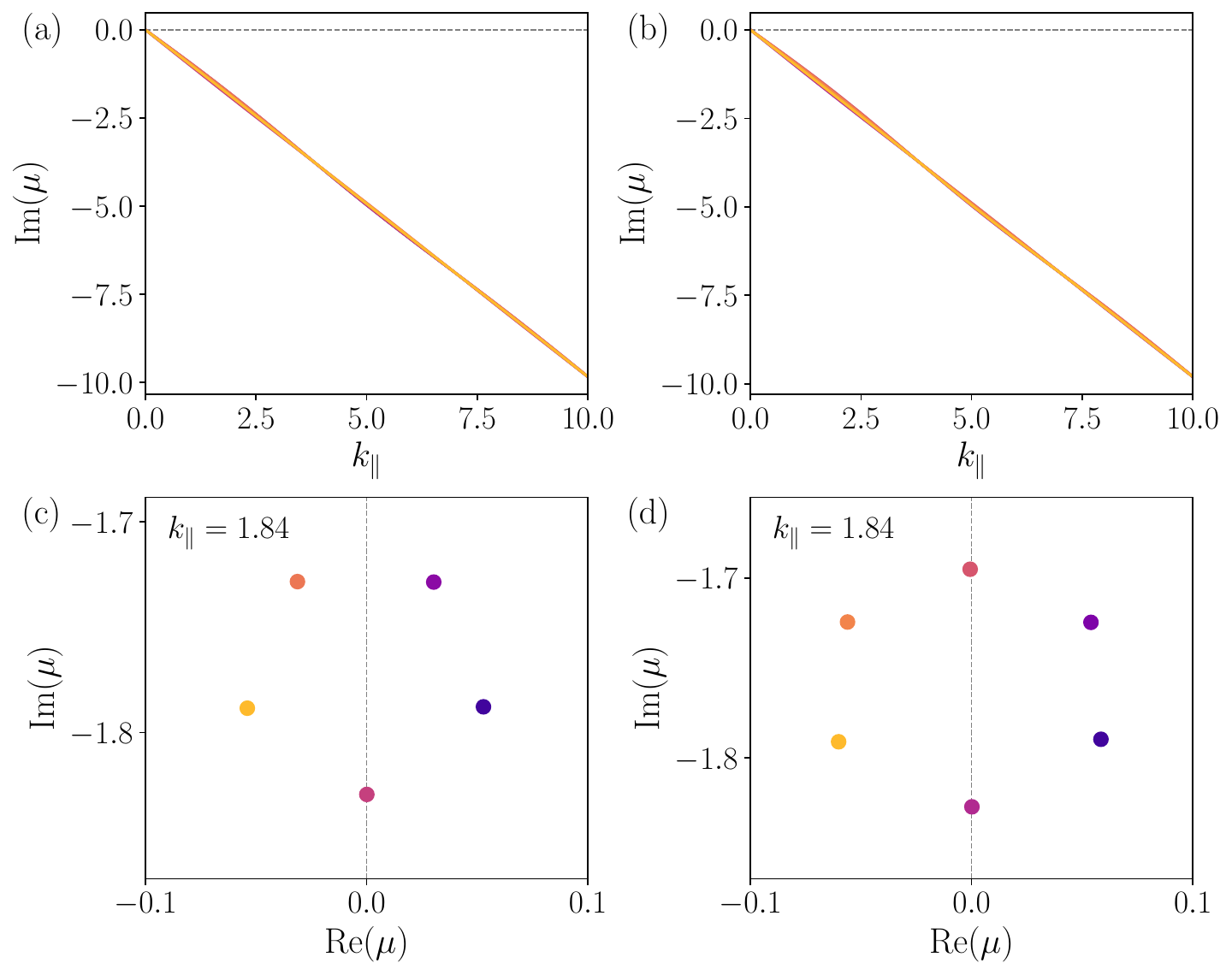}
	\caption{Dominant eigenvalues from linear stability around the ordered state in the $k_\parallel$ direction, for $m=5$ (left) and $m=6$ (right). (a), (b) Imaginary part $\Im(\mu)$ of the $m$ most dominant eigenvalues as a function of $k_\parallel$, colored from most to least dominant at $k_\parallel=1.84$. (c), (d) The same dominant eigenvalues plotted in the complex plane at $k_\parallel=1.84$. Parameters correspond to the growth rates shown in Figs.~\ref{fig:multipop_phase_diagram}(c) and (f): $\gAA=-0.2$, $\gAB=0.3$, $\rho=100$, $D_r=0.2$ and $N_c=50$.}
	\label{fig:multi_stripes_eigs_im}
\end{figure}

For the instability around the ordered state for the alternatively ordered flocking stripes in Figs.~\ref{fig:multipop_phase_diagram}(c) and (f), the relevant eigenvalues always have $\Im(\mu)<0$, as shown in Fig.~\ref{fig:multi_stripes_eigs_im}, so the eigenvector coefficients are read clockwise, giving a predicted stripe ordering of $s \to s\pm 1$, depending on which instability peak is examined.

\clearpage


%

\end{document}